\documentclass[11pt]{article}

\usepackage{amsthm}
\usepackage{amsmath}
\usepackage{amssymb}
\usepackage{bm} 
\usepackage{bbm}
\usepackage{mathrsfs}
\usepackage{mathtools}
\usepackage{dsfont}
\usepackage{float}
\usepackage{bm}
\usepackage{natbib}
\setcitestyle{authoryear,open={(},close={)}}

\usepackage{graphicx}
\usepackage{adjustbox}
\usepackage{booktabs}
\usepackage{multirow}
\usepackage{caption}
\usepackage[dvipsnames]{xcolor}
\usepackage{comment}
\usepackage{nicefrac}
\usepackage{enumitem}
\usepackage{apptools}
\AtAppendix{\counterwithin{lemma}{section}}
\AtAppendix{\counterwithin{corollary}{section}}
\AtAppendix{\counterwithin{proposition}{section}}
\AtAppendix{\counterwithin{theorem}{section}}
\AtAppendix{\counterwithin{figure}{section}}
\AtAppendix{\counterwithin{table}{section}}
\AtAppendix{\counterwithin{remark}{section}}

\usepackage{subcaption}
\numberwithin{equation}{section}
\theoremstyle{plain}

\theoremstyle{definition}

\usepackage{xspace}
\usepackage[nolist,smaller]{acronym}  
\PassOptionsToPackage{hyphens}{url}\usepackage[colorlinks=true, allcolors=blue]{hyperref}

\newcommand{\acro}[1]{\textsc{#1}\xspace }

\usepackage{xspace}
\usepackage[nolist,smaller]{acronym}  

\PassOptionsToPackage{hyphens}{url}\usepackage[colorlinks=true, allcolors=blue]{hyperref}

\newcommand{\HMM}{\acro{\smaller HMM}}%Hiden Markov Model
\newcommand{\HMMs}{\acro{\smaller HMM\footnotesize{s}}}%Hiden Markov Models
\newcommand{\HSMM}{\acro{\smaller HSMM}}%Hidden Semi Markov Model
\newcommand{\HSMMs}{\acro{\smaller HSMM\footnotesize{s}}}%Hiden semi Markov Models
\newcommand{\MC}{\acro{\smaller MC}}%Monte Carlo 
\newcommand{\MCMC}{\acro{\smaller MCMC}}%Markov Chain  Monte carlo 
\newcommand{\HMC}{\acro{\smaller HMC}}%Hamiltonian Monte Carlo
\newcommand{\NUTS}{\acro{\smaller NUTS}}%No U-Turn Sampler

\newcommand{\EM}{\acro{\smaller EM}} %EM
\newcommand{\MLE}{\acro{\smaller MLE}} %EM
\newcommand{\MSE}{\acro{\smaller MSE}} %EM
\newcommand{\ESS}{\acro{\smaller ESS}} %EM

\newcommand{\PA}{\acro{\smaller PA}} 
\newcommand{\IA}{\acro{\smaller IA}} 
\newcommand{\MA}{\acro{\smaller MA}} 
\newcommand{\HA}{\acro{\smaller HA}} 

\newcommand{\AIC}{\acro{\smaller AIC}} 
\newcommand{\BIC}{\acro{\smaller BIC}}

\newcommand{\stan}{\textit{stan}}

\DeclareMathOperator*{\argmax}{arg\,max} % Jan Hlavacek

\def\@fnsymbol#1{\ensuremath{\ifcase#1\or *\or \dagger\or \ddagger\or
   \mathsection\or \mathparagraph\or \|\or **\or \dagger\dagger
   \or \ddagger\ddagger \else\@ctrerr\fi}}
\newcommand{\ssymbol}[1]{^{\@fnsymbol{#1}}}

\let\OLDthebibliography\thebibliography
\renewcommand\thebibliography[1]{
  \OLDthebibliography{#1}
  \setlength{\parskip}{0pt}
  \setlength{\itemsep}{3pt plus 0.3ex}
}

\def\*#1{\bm{#1}} %\def\*#1{#1}
\usepackage[left=2.3cm, right=2.3cm, top=3cm]{geometry}
\usepackage{authblk}
\title{\bf  {Bayesian Approximations to  Hidden Semi-Markov Models for Telemetric Monitoring of Physical Activity }}

\author[1]{Beniamino Hadj-Amar}
\author[2]{Jack Jewson}
\author[3]{Mark Fiecas}
\affil[1]{Department of Statistics, Rice University, TX 77005-1827}
\affil[2]{Barcelona Graduate School of Economics, Universitat Pompeu Fabra, Spain, 08005}
\affil[3]{Division of Biostatistics, School of Public Health,
	University of Minnesota, Minneapolis, MN 55455}
\affil[ ]{\textit {\textcolor{blue}{beniamino.hadj-amar@rice.edu, jack.jewson@upf.edu, mfiecas@umn.edu}}}%, 
\date{May 2022}

\begin{document}

%\doparttoc % Tell to minitoc to generate a toc for the parts
%\faketableofcontents % Run a fake tableofcontents command for the partocs

%\part{} % Start the document part
%\parttoc % Insert the document TOC

%\bibliographystyle{natbib}

\def\spacingset#1{\renewcommand{\baselinestretch}%
{#1}\small\normalsize} \spacingset{1}

\setcounter{Maxaffil}{0}
\renewcommand\Affilfont{\itshape\small}

\spacingset{1.42} % DON'T change the spacing!

\maketitle
\begin{abstract}
We propose a Bayesian hidden Markov model for analyzing time series and sequential data where a special structure of the transition probability matrix is embedded to model explicit-duration semi-Markovian dynamics. Our formulation allows for the development of highly flexible and interpretable models that can integrate available prior information on state durations while keeping a moderate computational cost to perform efficient posterior inference. We show the benefits of choosing a Bayesian approach for \HSMM estimation  over its frequentist counterpart, in terms of 
%incorporation of  prior information, quantification of uncertainty
model selection and out-of-sample forecasting, also highlighting the computational feasibility of our inference procedure whilst incurring negligible statistical error. The use of our methodology is illustrated in an application relevant to e-Health, where we investigate rest-activity rhythms using telemetric activity data collected via a wearable sensing device. This analysis considers for the first time Bayesian model selection for the form of the explicit state dwell distribution. We further investigate the inclusion of a circadian covariate into the emission density and estimate this in a data-driven manner.
%We illustrate the use of our methodology on e-Health data, where we investigate rest-activity rhythms by analyzing physical activity time series collected via a wearable sensing device.

% \color{red} Our analysis of this data considers for the first time Bayesian model selection for the form of the explicit state dwell distribution and includes a circadian covariate, estimated in a data-driven manner, into the emission distribution. \color{black}
\end{abstract}

\noindent % 
\spacingset{1.45} % DON'T change the spacing!

\section{Introduction}

Recent developments in portable computing technology and the increased popularity of wearable and non-intrusive devices, e.g. smartwatches, bracelets, and smartphones, have provided exciting opportunities to measure and quantify physiological time series that are of interest in many applications, including mobile health monitoring, chronotherapeutic healthcare and cognitive-behavioral treatment of insomnia \citep{williams2013cognitive,kaur2013timing,silva2015mobile,aung2017sensing, huang2018hidden}.  The behavioral pattern of alternating sleep and wakefulness in humans can be investigated by measuring gross motor activity.  Over the last twenty years, activity-based sleep-wake monitoring has become an important assessment tool for quantifying the quality of sleep \citep{ancoli2003role,sadeh2011role}\color{black}. Though polysomnography \citep{douglas1992clinical}, usually carried out within a hospital or at a sleep center, continues to remain the gold standard for diagnosing sleeping disorders, accelerometers  have
become a practical and inexpensive way to collect non-obtrusive and continuous measurements of rest-activity rhythms over a multitude of days in the individual’s home sleep environment \citep{ancoli2015sbsm}.

\begin{figure}[htbp] 
	\centering
	\centerline{\includegraphics[width = \columnwidth]{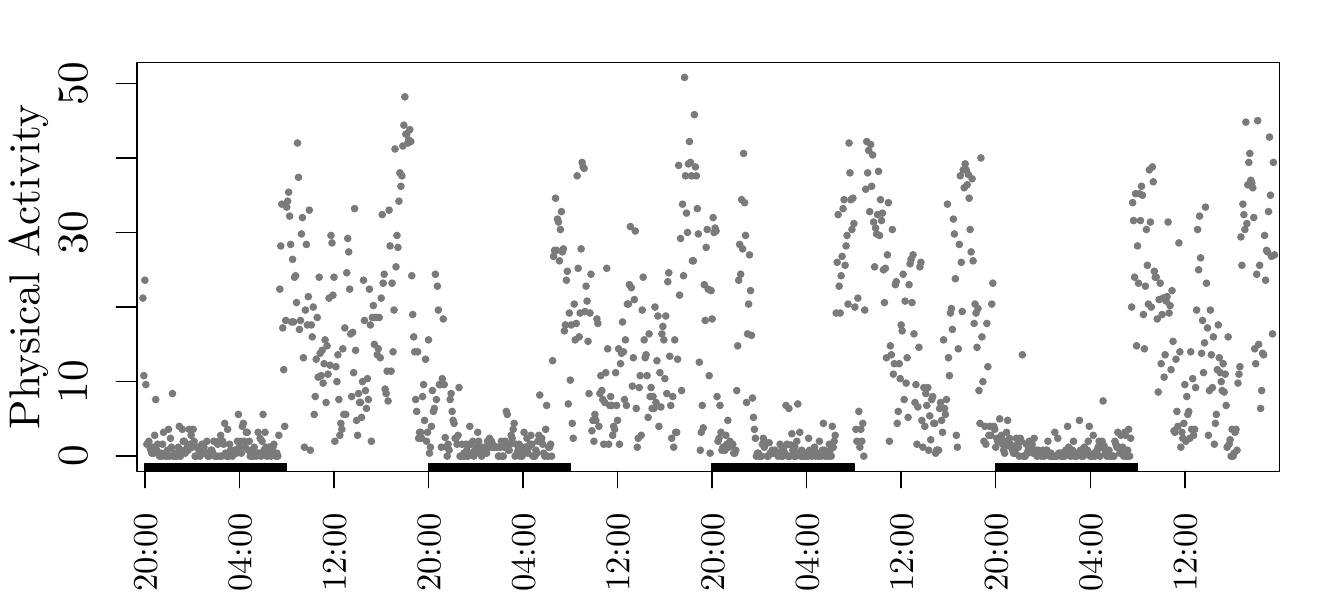}}
	\caption{\PA time series for a healthy individual. Rectangles on the time axis correspond to periods from 20.00 to 8.00.}
	\label{fig:physical_activity}
\end{figure}

Our study investigates the  \textit{physical activity} (\PA) time-series first considered by
\citet{huang2018hidden} and \citet{hadj2019bayesian}, where a wearable sensing device is fixed to the chest of a user to measure its movement via a triaxial accelerometer (ADXL345, Analog Devices). The tool produces \PA counts, defined as the number of times an accelerometer undulation exceeds zero over a specified time interval. Figure \ref{fig:physical_activity} displays an example of 4 days of 5-min averaged \PA recordings for a healthy subject, providing a total of 1150 data points. Transcribing information from such complex, high-frequency data into interpretable and meaningful statistics is a non-trivial challenge, and there is a need for a data-driven procedure to automate the analysis of these types of measurements. %While \citet{huang2018hidden} addressed this task by proposing an \HMM within a frequentist framework, we formulate a Bayesian hybrid model that takes into account semi-Markovian dynamics and incorporates available prior information for different activity patterns.
While \citet{huang2018hidden} addressed this task by proposing a hidden Markov model (\HMM) within a frequentist framework, we formulate a more flexible approximate hidden semi-Markov model (\HSMM) approach that enables us to explicitly model the dwell time spent in each state. Our proposed modelling approach uses a Bayesian inference paradigm, allowing us to incorporate available prior information for different activity patterns and facilitate consistent and efficient model selection between dwell distribution.

%\begin{figure}[htbp] 
%	\centering
%	\centerline{\includegraphics[height = 6cm, width = 13.5cm]{Plots/physical_activity.png}}
%	\caption{\jack{Update to .pdf} \PA time series for a healthy individual. Rectangles on the time axis correspond to periods from 20.00 to 8.00.}
%	\label{fig:physical_activity}
%\end{figure}

%We propose a Bayesian \HMM  that is a reformulation of any given \HSMM. We 
We conduct Bayesian inference using a \HMM likelihood model that is a reformulation of any given \HSMM. We 
utilise the method of \citet{langrock2011hidden} to embed the generic state duration distribution within a special transition matrix structure that can approximate the underlying \HSMM with arbitrary accuracy. %This framework enables the development of highly flexible and interpretable models that can incorporate available prior information on state durations,  without renouncing the well-developed methodology that characterizes \HMMs as a tractable structure for univariate and multivariate time series. \jack{Note about being able to explicitly model the dwell distribution, extra flexibility, and the convenience of \HMMs, we achieve a middle}  
This framework is able to incorporate the extra flexibility of explicitly modelling the state dwell distribution provided by a \HSMM, without renouncing the computational tractability, theoretical understanding, and the multitudes of methodological advancements that are available when using an \HMM.
To the best of our knowledge, such a modeling approach has only previously been treated from a non-Bayesian perspective in the literature, where parameters are estimated either by direct numerical likelihood maximization (\MLE) or applying the expectation-maximization (\EM) algorithm.

 The main practical advantages of a fully Bayesian framework for \HSMM inference 
%as opposed to \cite{langrock2011hidden} 
are that the regularisation and uncertainty quantification provided by the prior and posterior distributions can be readily incorporated into improved mechanisms for prediction and model selection. In particular, selecting the \HSMM dwell distribution in a data-driven manner and performing predictive inference for future state dwell times.

However, the posterior distribution is rarely available in closed form and the computational burden of approximating the posterior, often by sampling \citep[see e.g.][]{gelfand1990sampling}, is considered a major drawback of the Bayesian approach. 
In particular,  evaluating the likelihood in \HSMMs is already computationally burdensome \citep{guedon2003estimating}, yielding implementations that are often prohibitively slow. This further motivates the use of the likelihood approximation of \citet{langrock2011hidden} within a Bayesian framework. Here, we combine their approach with the \stan{} probabilistic programming language \citep{carpenter2016stan}, further accelerating the likelihood evaluations by proposing a sparse matrix implementation and leveraging \stan's compatibility with bridge sampling \citep{meng1996simulating,meng2002warp,gronau2017bridgesampling} to facilitate Bayesian model selection. We provide examples to illustrate the statistical advantages of our Bayesian implementation in terms of prior regularization, forecasting, and model selection and further illustrate that the combination of our approaches can %allow the computational feasibility of such inference (for example, by reducing the time for inference from more than three days to less than two hours), whilst incurring negligible statistical error.
make such inferences computationally feasible (for example, by reducing the time for inference from more than three days to less than two hours), whilst incurring negligible statistical error.

%\jack{Answers to Questions 4 and 5 to the reviewers. We are agreed that Bayes provides better inference. However, it is also almost always more computationally intensive, therefore Bayesian inference for HSMM, an already computationally expensive likelihood to compute, is plagued by computational difficulties, we leverage \cite{langrock2011hidden}'s likelihood, the \stan{} modelling language, and sparse matrix formulation to make this inference feasible. We further show the benefits such inferences have from a statistical perspective and compare them with the original \cite{langrock2011hidden} as well as a fully Bayesian implementation. Lastly, we do something cool with the applied example }

% We choose not to discuss the philosophical desirability of either paradigm, this has been considered at length before \jack{references}, instead, we focus on the practical implications of the difference in these approaches

The rest of this article is organized as follows. In Section  \ref{sec:hmm_hsmm}, we provide a brief introduction to \HMMs and \HSMMs. Section \ref{sec:model} reviews the \HSMM likelihood approximation of \cite{langrock2011hidden}. Section \ref{sec:bayesian_inference} presents our Bayesian framework and inference approach.  Using several simulation studies, Section \ref{sec:simulation_studies} investigates the performance of our proposed procedure when compared with the implementation of \cite{langrock2011hidden}. Section \ref{sec:approx_accuracy} evaluates the trade-off between computational efficiency and statistical accuracy of our method  and proposes an approach to investigate the quality of the likelihood approximation for given data.  Section \ref{sec:application} illustrates the use of our method to analyze telemetric activity data,  and we further investigate the inclusion of spectral information within the emission density in Section \ref{sec:harmonic_emissions}. The \stan{} files (and  \texttt{R} utilities) that were used to implement  our experiments are available at 
\url{https://github.com/Beniamino92/BayesianApproxHSMM}.  The probabilistic programming framework associated with \stan{} makes it easy for practitioners to consider further dwell/emission distributions to the ones considered in this paper. Users need only change the corresponding function in our \stan{} files. 
%\href{https://github.com/Beniamino92/BayesianApproxHSMM}{https://github.com/Beniamino92/BayesianApproxHSMM}.

% Additionally, the latter allows us to learn appropriate dwell distributions from the data.

\color{black}

\section{Modeling Approach} \label{sec:modeling_approach}

% \textcolor{black}{We provide a brief introduction to the standard \HMM and \HSMM approaches before considering the special structure of the transition matrix presented by \cite{zucchini2017hidden}, which allows the state dwell distribution to be generalized with arbitrary accuracy. Such a modeling approach embodies the flexibility of the \HSMM framework to explicitly model the state dwell distribution while still leveraging the computational tractability, theoretical support, and methodological developments fundamental to \HMMs. We consider such a methodology for the first time within a Bayesian paradigm and later demonstrate the advantages of such a modeling approach.}

% Such a modeling approach embodies the flexibility of the \HSMM framework to explicitly model the state dwell distribution while still leveraging the computational tractability, theoretical support, and methodological developments fundamental to \HMMs. 
% We consider such models for the first time within a Bayesian paradigm and later demonstrate the advantages of such an approach.

\subsection{ Overview of Hidden Markov and Semi-Markov Models} 
 \label{sec:hmm_hsmm} 
We now provide a brief introduction to the standard \HMM and \HSMM approaches before considering the special structure of the transition matrix presented by \cite{zucchini2017hidden}, which allows the state dwell distribution to be generalized with arbitrary accuracy. 
%  Such a modeling approach embodies the flexibility of the \HSMM framework to explicitly model the  state dwell distribution  while still leveraging the computational tractability, theoretical support and methodological developments fundamental to \HMMs. We consider such a methodology for the first time within a Bayesian paradigm and later demonstrate the advantages of such a modeling approach.}
 \HMMs, or Markov switching processes, have been shown to be appealing models in addressing learning challenges in time series data and have been successfully applied in fields such as speech recognition \citep{rabiner1989tutorial, jelinek1997statistical}, digit recognition \citep{raviv1967decision, rabiner1989high} as well as  biological and physiological data  \citep{ langrock2013combining, huang2018hidden, hadj2020spectral}. An \HMM is a stochastic process model based on an unobserved (hidden) state sequence $\bm{s} = (s_1, \dots, s_T)$ that takes discrete values in the set $\{1, \dots, K\}$ and whose transition probabilities follow a Markovian structure. Conditioned on this state sequence, the observations $\bm{y} = (y_1, \dots, y_T)$ are assumed to be  conditionally independent and  generated from a parametric family of probability distributions $f(\bm{\theta}_j)$,  which are often called \textit{emission} distributions. This generative process can be outlined as 
\begin{equation}
\begin{split}
s_{\,t} \, | \, s_{\,t-1} &\sim \bm{\gamma}_{s_{\, t-1}} \\
y_t \, | \,  s_{\,t} \, &\sim \, f \, ( \,  \bm{\theta}_{s_{\,t}}) \qquad t = 1, \dots, T,
\end{split}
\label{eq:HMM}
\end{equation}
where $\bm{\gamma}_{\, j} = (\gamma_{j1}, \dots,  \gamma_{jK})$ denotes the state-specific vector of transition probabilities, $\gamma_{jk} = p \, (\, s_t = k \, | \, s_{t-1}  = j)$ with $\sum_{k} \gamma_{jk} = 1$, and $p\,(\cdot)$ is a generic notation for probability density or mass function, whichever appropriate. The initial state $s_0$ has distribution $\bm{\gamma}_0 = (\gamma_{01}, \dots, \gamma_{0K})$ and  $\bm{\theta}_j$ represents the vector of emission parameters modelling state $j$.  \HMMs provide a simple and flexible mathematical framework that can be naturally used for many inference tasks, such as signal  extraction,  smoothing,  filtering \color{black} and forecasting  (see e.g.  \citealt{zucchini2017hidden}). These appealing features are a result of an extensive theoretical and methodological literature that includes  several dynamic programming algorithms  for computing the likelihood in a straightforward and inexpensive manner (e.g. forward messages scheme, \citealt{rabiner1989tutorial}).  \HMMs are also naturally suited for local and global decoding (e.g. Viterbi algorithm, \citealt{forney1973viterbi}), and the incorporation of trend, seasonality and covariate information in both the observed process and the latent sequence. Although computationally convenient, the Markovian structure of \HMMs limits their flexibility. In particular, the \textit{dwell} duration in any state, namely the number of consecutive time points that the Markov chain spends in that state, is implicitly forced to follow a geometric distribution with probability mass function $p_j (d) = (1 - \gamma_{jj}) \, \gamma_{jj}^{\,d - 1}$.

\begin{figure}[htbp]
	\centering
	\includegraphics[scale = 0.3]{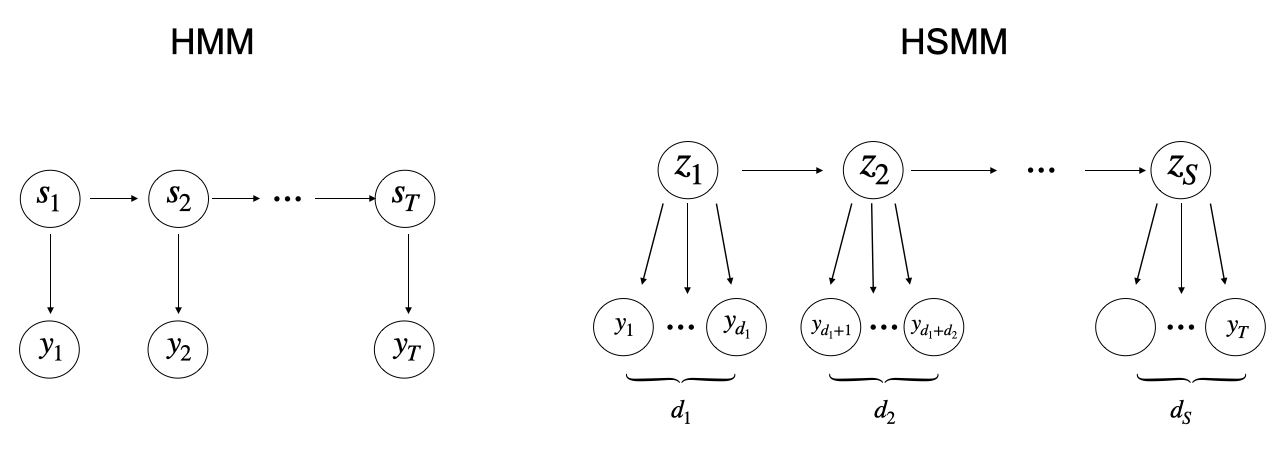}
	\caption{ Graphical models: (left) \HMM where $y_1, \dots, y_T$ are the observations and $s_1, \dots, s_T$ the corresponding hidden state sequence; (right) \HSMM where $d_1, \dots, d_S$ are the random dwell-times associated with each \textit{super} state of the Markov chain  $z_1, \dots, z_S$ where no self-transitions are allowed.}
	\label{fig:HMM_vs_HSMM}
\end{figure}

A more flexible framework can be formulated  using \HSMMs, where the generative process of an \HMM is augmented by introducing an explicit, state specific, form for the dwell time \citep{guedon2003estimating, johnson2013bayesian}. The state stays unchanged until the duration terminates, at which point there is a Markov transition to a new regime. As depicted in Figure \ref{fig:HMM_vs_HSMM}, the \textit{super-states}  $\bm{z} = (z_1, \dots, z_S)$ are generated from a Markov chain prohibiting self-transitions wherein each super-state $z_s$ is associated with a  dwell time $d_s$ and a random segment of observations $\bm{y}_s = (y_{t_s^1}, \dots, y_{t_s^2})$, where $t_s^1 = 1 +  \sum_{r<s} d_r$ and $t_s^2 = t_s^1 + d_s - 1$ represent the first and last index of segment $s$,  and $S$ is the (random) number of segments. Here, $d_s$ represents the length of the dwell duration of $z_s$. The generative mechanism of an \HSMM can be summarized as  \begin{equation}
\begin{split}
z_{\,s} \, | \, z_{\,s-1} &\sim \bm{\pi}_{\,z_{\, s-1}} \\ 
d_s \, | \, z_s \,  &\sim g \, ( \,  \bm{\lambda}_{\, z_s} ) \\ \bm{y}_{s}
\, | \,  z_{\,s} \, &\sim \, f \, (  \,  \bm{\theta}_{z_{\,s}}) \qquad s = 1, \dots, S, 
\end{split}
\label{eq:HSMM}
\end{equation}
where  $\bm{\pi}_{\, j} = (\pi_{j1}, \dots,  \pi_{jK})$ are state-specific transition probabilities in which  $\pi_{jk} = p \, (\, z_t = k \, | \, z_{t-1}  = j, \, z_t \neq j)$ for $j, k = 1, \dots, K$.  Note that $\pi_{jj} = 0$, since self transitions are prohibited.  We assume that the initial state has distribution $\bm{\pi}_0 = (\pi_{01}, \dots, \pi_{0K})$,  namely $\bm{z}_0 \sim \bm{\pi}_0$. Here, $g$ denotes a family of dwell distributions parameterized by some state-specific duration parameters $\bm{\lambda}_j$, which could be either a scalar (e.g. rate of a Poisson distribution), or a vector (e.g. rate and dispersion parameters for negative binomial durations). Unfortunately,  this increased flexibility in  modeling the state duration  has the cost of   substantially increasing the computational burden of computing the likelihood: the message-passing procedure for \HSMMs requires $\mathcal{O} \, (T^2 K + T K^2)$ basic computations for a time series of length $T$ and number of states $K$, whereas the corresponding forward-backward  algorithm  for \HMMs requires only $\mathcal{O} \, (T K^2)$.

\subsection{Approximations to Hidden Semi-Markov Models}
\label{sec:model}

%\jack{We could split this Section, separate Langrock and Zucchini as part of Section 2, and then talk about introducing the prior in Section 3 - answer R1 point 1. This is just an approximate likelihood evaluation rather than a different model}

In this section we introduce the \HSMM likelihood approximation of \citet{langrock2011hidden}.  Let us consider an \HMM in which $\bm{y}^{\star} = (y^{\star}_1, \dots, y^{\star}_T$) represents the observed process and $\bm{z}^{\star} = (z^{\star}_1 , . . . , z^{\star}_T )$ denotes the latent discrete-valued  sequence of a Markov chain with states $\{1, 2, \dots, \bar{A} \}$, where  $\bar{A} = \sum_{i = 1}^{K} a_i$, and $a_1, \dots, a_K$ are arbitrarily fixed positive integers.   Let us define \textit{state aggregates} $A_j$ as \begin{equation}
A_j = \Bigg\{ \,  a : \sum_{i=0}^{j-1} a_i < a \leq \sum_{i=0}^{j} a_i \,   \Bigg\}, \quad j = 1, \dots, K,
\end{equation} where $a_0 = 0$, and each state corresponding to $A_j$ is associated with the same emission distribution $f ( \, \bm{\theta}_j )$ in the  \HSMM formulation  of Eq. \eqref{eq:HSMM}, namely $    y^{\star}_t \, \big| \,  z^{\star}_{\,t} \in A_j  \sim f \, ( \bm{\theta}_j).$ The probabilistic rules governing the transitions between states $\bm{z}^\star$ are described via the matrix $\bm{\Phi} = \big\{  \phi_{il} \big\}$, where $\phi_{il} = p \, (\, z^\star_{\, t} = l \, | \, z^\star_{\, t-1}  = i \, )$, for $i, l = 1, \dots, \bar{A}$. This matrix has the following structure

\begin{equation}
\label{eq:phi_mat}
\bm{\Phi} = \begin{bmatrix} 
\bm{\Phi}_{11} & \dots & \bm{\Phi}_{1K} \\
\vdots & \ddots & \vdots \\
\bm{\Phi}_{K1} & \dots       & \bm{\Phi}_{KK} 
\end{bmatrix},
\end{equation} where the sub-matrices $\bm{\Phi}_{jj}$ along the main diagonal, of dimension $a_j \times a_j $, are defined for $a_j \geq 2$, as \begin{equation}
\bm{\Phi}_{jj} = 
\begin{bmatrix}
0 & 1 - h_j\,(1) & 0 & \dots & 0 \\
\vdots & 0 & \ddots &  & \vdots \\
& \vdots &  &  & 0 \\
0 & 0 & \dots & 0 & 1 - h_j\,(a_j - 1) \\
0 & 0 & \dots & 0 &  1 - h_j\,(a_j)
\end{bmatrix},
\end{equation}
and $\bm{\Phi}_{jj} = 1 - h_j(1)$, for $a_j = 1$. The $a_j \times a_k$ off-diagonal matrices $\bm{\Phi}_{jk}$ are given by 
\begin{equation}
\bm{\Phi}_{jk} = 
\begin{bmatrix}
\pi_{jk}\, h_j\,(1) & 0 & \dots & 0 \\
\pi_{jk}\, h_j\,(2) & 0 & \dots & 0 \\
\vdots &  &  & \\
\pi_{jk} \, h_j\,(a_j) & 0 & \dots & 0
\end{bmatrix}
\end{equation}
where in the case that $a_j = 1$ only the first  column is included. Here, $\pi_{jk}$ are the transition probabilities of an \HSMM as in Eq. \eqref{eq:HSMM}, and the \textit{hazard rates} $h_j \, (r)$ are specified for $ r \in \mathbb{N}_{> 0} $ as
\begin{equation}
h_j \, (r) = \dfrac{p \, (\, d_j = r \, | \,  \bm{\lambda}_j)}{p \, (\, d_j \geq r \, | \,  \bm{\lambda}_j)}, \quad \text{if} \, \, p \, (\, d_j \geq r -1  \, | \,  \bm{\lambda}_j) < 1,
\label{eq:hazard_rates}
\end{equation}
and 1 otherwise, where $p \, (\, d_j = r \, | \,  \bm{\lambda}_j)$  denotes the probability mass function of the dwell distribution $g \, ( \bm{\lambda}_j)$ for state $j$. This structure for the matrix $\bm{\Phi}$ implies that transitions within state aggregate $A_j$ are determined by diagonal matrices $\bm{\Phi}_{jj}$, while transitions between state aggregates $A_j$ and $A_k$ are controlled by off-diagonal matrices $\bm{\Phi}_{jk}$. Additionally, a transition from $A_j$ to $A_k$ must enter $A_k$ in $\min(A_k)$. \citet{langrock2011hidden} showed that this choice of $\bm{\Phi}$  allows for the representation of any duration distribution, and yields an \HMM that is, at least approximately, a reformulation of the underlying \HSMM. \color{black} In summary, the distribution of $\bm{y}$  (generated from an underlying \HSMM) \color{black} can be approximated by that of $\bm{y}^\star$  (modelled using $\bm{\Phi}$)\color{black}, and this approximation can be designed to be arbitrarily accurate by choosing $a_j$ adequately large. In fact, the representation of the dwell distribution through $\bm{\Phi}$  differs from the true distribution, namely the one in the \HSMM formulation of Eq. \eqref{eq:HSMM}, only for values larger than $a_j$, i.e., in the right tail. %\color{red} For this reason we term $a$ the \textit{dwell thresholds}.\color{black}

\section{Bayesian Inference}
\label{sec:bayesian_inference}

\begin{figure}[htbp] 
	\centering
	\includegraphics[scale = 0.18]{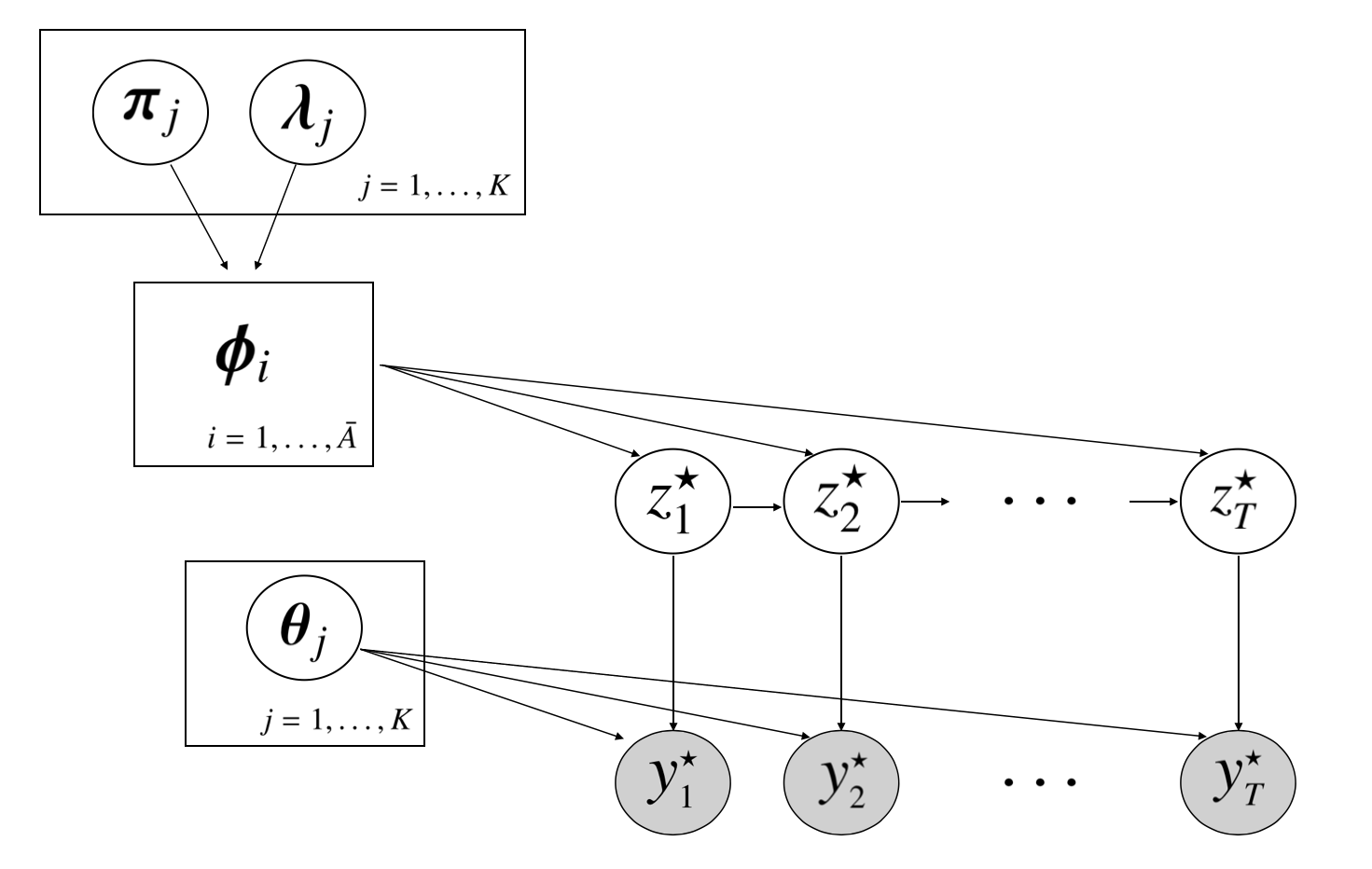}
	\caption{A graphical model for Eq. \eqref{eq:bayes_model}. Transition probabilities $\bm{\phi}_j$ are solely determined by $\bm{\pi}_j$ and $p \, (\, d_j = r \, | \,  \bm{\lambda}_j)$, and thus they are not considered as random variables themselves.}
	\label{fig:bayes_graphical_model}
\end{figure}

Bayesian inference for \HSMMs has long been plagued by the computational demands of evaluating its likelihood. In this section we use the \HSMM likelihood approximation of \cite{langrock2011hidden} to facilitate efficient Bayesian inference for \HSMMs. Extending the model introduced in Section \ref{sec:model} to the Bayesian paradigm requires placing priors on the model parameters $\bm{\eta} = \big\{ \, (  \bm{\pi}_j, \, \bm{\lambda}_j, \, \bm{\theta}_j ) \, \big\}_{j=1}^{\,K}$. 
%
%\jack{Move this to Section 3 to start talking about Bayes. We propose to make Bayesian inference feasible for \HSMMs by using Langrock and Zucchini's likelihood evaluation within a Bayesian paradigm. We do this because Bayes is useful for HSMMs and the approximation mixed with stan makes this feasible }
%
The generative process of our Bayesian model can be summarized  by \begin{equation}
\begin{split}
\bm{\pi}_j &\sim \text{Dir}\, (\bm{\alpha}_0), \qquad (\bm{\theta}_j, \bm{\lambda}_j ) \sim H \times G, \qquad j = 1, \dots, K, \\
z^\star_{\,t} \, | \, z^\star_{\,t-1} &\sim \bm{\phi}_{\,z^\star_{\, t-1}} \\
\bm{y}^\star_{t}
\, | \,  z^\star_{\,t} \in A_j \, &\sim \, f \, (  \,  \bm{\theta}_{j})  \, \hspace{4.6cm}  t = 1, \dots, T,
\end{split}
\label{eq:bayes_model}
\end{equation} 
where Dir$(\cdot)$ denotes the Dirichlet distribution over a $(K-2)$ dimensional simplex (since the probability of self transition is forced to be zero) and $\bm{\alpha}_0$ is a vector of positive reals. Here, $H$ and $G$ represent the priors over emission and duration parameters, respectively, and $\bm{\phi}_i$ denotes the $i^{th}$ row of the matrix $\bm{\Phi}$. A graphical model representing the probabilistic structure of our approach is shown in Figure \ref{fig:bayes_graphical_model}, where we remark  that the entries of the transition  matrix $\bm{\Phi}$ are entirely determined by the transition probabilities of the Markov chain $\bm{\pi}_{j}$ and the values of the durations $p \, (\, d_j = r \, | \,  \bm{\lambda}_j)$.

%\subsection{Inference} 
%\label{sec:bayesian_inference}
%Our inference scheme for estimating the  parameters $\bm{\eta} = \big\{ \, (  \bm{\pi}_j, \, \bm{\lambda}_j, \, \bm{\theta}_j ) \, \big\}_{j=1}^{\,K}$ is  formulated within a full Bayesian framework  and it is based on the following factorization of the joint posterior distribution 
The posterior distribution for $\bm{\eta}$ has the following factorisation.
\begin{equation}
p \, ( \,  \bm{\eta} \, | \, \bm{y} ) \propto \mathscr{L}\,  ( \bm{y} \, | \,  \bm{\eta} ) \, \times \, \bigg[ \, \prod_{j=1}^{K} \,  p \, ( \bm{\pi}_j ) \, \times \,   p \, ( \bm{\lambda}_j  ) \, \times \, p \, ( \bm{\theta}_j) \, \bigg] \, ,
\label{eq:posterior}
\end{equation} where $\mathscr{L}\, ( \, \cdot \, )$ denotes the likelihood of the model, $p \, ( \bm{\pi}_j)$ is the density of the Dirichlet prior for  transitions probabilities (Eq. \ref{sec:model}), and $p \, ( \bm{\lambda}_j)$ and $p \, ( \bm{\theta}_j)$ represent the prior densities for dwell and emission parameters, respectively. Since we have
formulated an \HMM, we can employ well-known techniques that are available to compute the likelihood, and in particular we can express it using the following matrix multiplication (see e.g. \citealt{zucchini2017hidden})
\begin{equation}
\mathscr{L}\,  ( \bm{y} \, | \bm{\eta} ) =  \bm{\pi}_0^{\,\star\,'} \, \bm{P}\,(y_1) \,  \bm{\Phi} \,  \bm{P}\,(y_2) \, \bm{\Phi} \, \cdots \, \bm{\Phi} \, \bm{P}\,(y_{T-1}) \,  \bm{\Phi} \, \bm{P}\,(y_{T}) \, \mathbf{1},
\label{eq:likelik}
\end{equation} where the diagonal matrix $\bm{P}\,(\,y\,)$ of dimension $\bar{A} \times \bar{A}$ is defined as \begin{equation}
\bm{P}\,(\,y\,) =  \text{diag} \, \big\{ \, \underbrace{p \,( y \, | \, \bm{\theta}_1), \,  \dots,\, p  \,( y \, | \, \bm{\theta}_1)}_{a_1 \, \, \text{times}}, \, \dots, \, \underbrace{p \, ( y \, | \, \bm{\theta}_K) \dots p \, ( y \, | \, \bm{\theta}_K)}_{a_K \, \, \text{times}} \big\},
\label{eq:diag_mat}
\end{equation}
and $ p \, ( y \, | \, \bm{\theta}_j)$ is the probability density of the emission distribution $f \, (  \,  \bm{\theta}_{j})$. Here, $\mathbf{1}$ denotes  an $\bar{A}$-dimensional column vector with all entries equal to one  and $\bm{\pi}_0^{\,\star}$ represents the initial distribution for the state aggregates.  Note that if we assume that the underlying Markov chain is stationary, $\bm{\pi}_0^{\,\star}$ is solely determined by the transition probabilities $\bm{\Phi}$,  i.e. $\bm{\pi}_0^{\,\star} \, =  ( \bm{I} - \bm{\Phi} + \bm{U})^{\,-1} \, \mathbf{1}$, where $\bm{I}$ is the identity matrix and $\bm{U}$ is a square matrix of ones.  Alternatively, it is possible to start from a specified state, namely assuming that $\bm{\pi}_0^{\,\star}$ is an appropriate unit vector, e.g. $(1, 0,  \dots, 0)$, as suggested by \citet{leroux1992maximum}. We finally note that computation of the likelihood in Eq. \eqref{eq:likelik} is often subject to numerical underflow and hence its practical implementation usually require appropriate scaling \citep{zucchini2017hidden}.

While a fully Bayesian framework is desirable for its ability to provide coherent uncertainty quantification for parameter values, a perceived drawback of this approach compared with a frequentist analogue is the increased computation required for estimation. Bayesian posterior distributions are only available in closed form under the very restrictive setting when the likelihood and prior are conjugate. Unfortunately, the model outlined in Section \ref{sec:model} does not admit such a conjugate prior form and as a result the corresponding posterior (Eq. \ref{eq:posterior}) is not analytically tractable. However, numerical methods such as Markov Chain Monte Carlo (\MCMC) can be employed to sample from this intractable posterior. The last twenty years have seen an explosion of research into \MCMC methods and more recently approaches scaling them to high dimensional parameter spaces. The next section outlines one such black box implementation that is used to sample from the posterior in Eq. \eqref{eq:posterior}.

\subsection{Hamiltonian Monte Carlo, No-U-Turn Sampler and Stan Modelling Language}{\label{sec:bayes_computation}}

One particularly successful posterior sampling algorithm is Hamiltonian Monte Carlo (\HMC, \citealt{duane1987hybrid}), where we refer the reader to \citet{neal2011mcmc} for an excellent introduction. \HMC augments the parameter space with a `momentum variable' and uses Hamiltonian dynamics to propose new samples. The gradient information contained within the Hamiltonian dynamics allows \HMC to produce proposals that can traverse high dimensional spaces more efficiently than standard random walk \MCMC algorithms. However, the performance of \HMC samplers is dependent on the tuning of the leapfrog discretisation of the Hamiltonian dynamics. The No-U-Turn Sampler (\NUTS) \citep{hoffman2014no} circumvents this burden. \NUTS uses the Hamiltonian dynamics to construct trajectories that move away from the current value of the sampler until they make a `U-Turn' and start coming back, 
% towards the original point,
thus maximising the trajectory distance. An iterative algorithm allows the trajectories to be constructed both forwards and backwards in time, preserving time reversibility. Combined with a stochastic optimisation of the step size, \NUTS is able to conduct efficient sampling without any hand-tuning.

%Does stan use the Riemanian geometry (second order infor also) as well?

The \stan{} modelling language \citep{carpenter2016stan} provides a probabilistic programming environment facilitating the easy implementation of \NUTS. The user needs only define the three components of their model: (i) the inputs to their sampler, e.g. data and prior hyperparameters; (ii) the outputs,  e.g. parameters of interest; (iii) the computation required to calculate the unnormalized posterior. Following this, \stan{} uses automatic differentiation \citep{griewank2008evaluating} to produce fast and accurate samples from the target posterior. \stan's easy-to-use interface and lack of required tuning have seen it implemented in many areas of statistical science.  %(\jack{references}). 
As well as using \NUTS to automatically tune the sampler, \stan{} is equipped with a variety of warnings and tools to help users diagnose the performance of their sampler. For example,  convergence of all quantities of interest is monitored in an automated fashion by comparing variation between and within simulated samples initialized at over-dispersed starting values \citep{gelman2017prior}. Additionally, the structure of the transition matrix $\bm{\Phi}$ allows us to take advantage of \stan's sparse matrix implementation to achieve vast computational improvements. Although $\bm{\Phi}$ has dimension $\bar{A} \times \bar{A}$, each row has at most $K$ non-zero terms (representing within state transitions to the next state aggregate or between state transitions),
% and as a result  $\bm{\Phi}$  \textcolor{red}{shows a proportion  $(K/\bar{A})$ of non zero elements}. 
 and as a result only a proportion $(K/\bar{A})$ of the elements of $\bm{\Phi}$ is non-zero. 
Hence, for large values of the dwell approximation thresholds $\bm{a}$, the matrix $\bm{\Phi}$ exhibits considerable sparsity. The \stan{} modelling language implements compressed row storage sparse matrix representation and multiplication, which provides considerable speed up when the sparsity is greater than 90\% \citep[][Ch. 6]{stan2018stan}. In our applied scenario we consider dwell-approximation thresholds as big as $\bm{a} = (250, 50, 50)$ with sparsity of greater than 99\% allowing us to take considerable advantage of this formulation.  Finally, we note that our proposed Bayesian approach may suffer from \textit{label switching} \citep{stephens2000dealing} since the likelihood is invariant under permutations of the labels of the hidden states.  However, this issue is easily addressed using order constraints provided by \stan. This strategy worked well in the simulations and applications presented in the paper, without introducing any noticeable bias in the results.

\subsection{Bridge Sampling Estimation of the Marginal Likelihood}
The Bayesian paradigm provides a natural framework for selecting between competing models by means of the marginal likelihood, i.e.
\begin{equation}
		    p \, (\bm{y}) = \int  \mathscr{L}\,  ( \bm{y} \, | \,  \bm{\eta} ) \, p \, (\bm{\eta}) \, d \bm{\eta}.
		    \label{eq:marg_likelik}
		\end{equation}
The ratio of marginal likelihoods from two different models, often called the \textit{Bayes factor} \citep{kass1995bayes}, can be thought of as the weight of evidence in favor of a model against a competing one. The marginal likelihood in Eq. \ref{eq:marg_likelik} corresponds to the normalizer of the posterior $p \,(  \bm{\eta} \, | \, \bm{y} )$ (Eq. \ref{eq:posterior}) and is generally the component that makes the posterior analytically intractable. \MCMC algorithms, such as the \stan's implementation of \NUTS introduced above, allow for sampling from the unnormalized posterior, but further work is required to estimate the normalizing constant. Bridge sampling \citep{meng1996simulating,meng2002warp} provides a general procedure for estimating these marginal likelihoods reliably. While standard Monte Carlo (\MC) estimates draw samples from a single distribution, bridge sampling formulates an estimate of the marginal likelihood using the ratio of two \MC estimates drawn from different distributions: one being the posterior (which has already been sampled from) and the other being an appropriately chosen proposal distribution $q\,(\bm{\eta})$. The bridge sampling estimate of the marginal likelihood is then given by 
\begin{equation*}
 p \, (\bm{y}) = \frac{\mathbb{E}_{\,q(\bm{\eta})}\left[h(\bm{\eta}) \, \mathscr{L}\,  ( \bm{y} \, | \,  \bm{\eta} ) \, p \, (\bm{\eta}) \, \right]}{\mathbb{E}_{\, p(\bm{\eta} | \bm{y})}\left[h(\bm{\eta})\,q(
 \bm{\eta})\right]} \approx \frac{\frac{1}{n_2}\sum_{j=1}^{n_2}h(\bm{\tilde{\eta}}^{\,(j)}) \, \mathscr{L}\,  ( \bm{y} \, | \,  \bm{\tilde{\eta}}^{\,(j)}) \, p \, (\bm{\tilde{\eta}}^{\,(j)})}{\frac{1}{n_1}\sum_{i=1}^{n_1}h\,(\bm{\overline{\eta}}^{\,(i)}) \, q(\bm{\overline{\eta}}^{\,(i)})}, \end{equation*}
where $h(\bm{\eta})$ is an appropriately selected \textit{bridge function} and $p (\bm{\eta})$ denotes the joint prior distribution. Here, $\{\bm{\overline{\eta}}^{\,(1)}, \ldots, \bm{\overline{\eta}}^{\,(n_1)}\}$ and $\{\bm{\tilde{\eta}}^{\,(1)}, \ldots, \bm{\tilde{\eta}}^{\,(n_2)}\}$ represent $n_1$ and $n_2$ samples drawn from the posterior $p \,(  \bm{\eta} \, | \, \bm{y} )$ and the proposal distribution $q(\bm{\eta})$, respectively. This estimator can be implemented in \texttt{R} using the package \texttt{bridgesampling} \citep{gronau2017bridgesampling}, whose compatibility with \stan{} makes it particularly straightforward to estimate the marginal likelihood directly from a \stan{} output. This package implements  the method of \cite{meng1996simulating} to choose the optimal bridge function minimising the estimator mean-squared error and constructs a  multivariate normal proposal distribution whose mean and variance match those of the sample from the posterior.

\subsection{Comparable Dwell Priors} \label{sec:comparable_dwell}

Model selection based on marginal likelihoods can be very sensitive to prior specifications. In fact, Bayes factors are only defined when the marginal likelihood under each competing model is proper \citep{robert2007bayesian, gelman2013bayesian}. As a result, it is important to include any available prior information into the Bayesian modelling in order to use these quantities in a credible manner.  Reliably characterising the prior for the dwell distributions is particularly important for the experiments considered in Section \ref{sec:application}, since we use Bayesian marginal likelihoods to select between the dwell distributions associated with \HSMMs and \HMMs. For instance, if we believe that the length of sleep for an average person is between 7 and 8 hours we would choose a prior that reflects those beliefs in all competing models. However, we need to ensure that we encode this information in \textit{comparable priors} in order to perform `fair' Bayes factor selection amongst a set of dwell-distributions. Our aim is to infer which dwell distribution, and not which prior specification, is most appropriate for the data at hand.

%We will be using Bayes factors to select between dwell distributions and thus where possible we seek to place comparable priors for the parameters of the dwell distributions for the HMM and HSMM{}s

% While we consider time-series with thousands of observations, long dwell times in any of the states result in there being relatively few state transitions and thus we expect the prior influence to be high. In particular, when modeling physical activity data in Section \ref{sec:application}, we focus on using Bayesian marginal likelihoods to select between the dwell distributions associated with \HSMMs and \HMMs.

For example, suppose we consider selecting between geometric (i.e. an \HMM), negative binomial or Poisson distributions (i.e. an \HSMM), to model the dwell durations of our data. While a Poisson random variable, shifted away from zero to consider strictly positive dwells, has its mean $\lambda_j + 1$ and variance $\lambda_j$ described by the same parameter $\lambda_j$, the negative binomial allows for further modelling of the precision through an additional factor $\rho_j$. In both negative binomial and Poisson \HSMMs, the parameters $\lambda_j$ are usually assigned a prior $\lambda_j \sim \text{Gamma}\,(a_{0j}, b_{0j})$ with mean $\mathbb{E}\left[\lambda_j\right] = a_{0j}/b_{0j}$ and variance $\textrm{Var}  \left[\lambda_j\right] = a_{0j}/b_{0j}^2$. In order to develop an interpretable comparison of all competing models,  we parameterize the geometric dwell distribution associated with state $j$ in the standard \HMM (Eq. \ref{eq:HMM}) as also being characterized by the mean dwell length $\tau_j = 1/(1-\gamma_{jj})$, where the geometric is also shifted to only consider strictly positive support and $\gamma_{jj}$ represents the probability of self-transition.  Under a Dirichlet prior for the state-specific vector of transition probabilities  $\bm{\gamma}_j = (\gamma_{j1}, \ldots, \gamma_{jK}) \sim \text{Dirichlet}(\bm{v}_j)$, with $\bm{v}_j = (v_{j1}, \ldots, v_{jK})$ and $\beta_j = \sum_{i\neq j} v_{ji}$, the mean and variance of the prior mean dwell under an \HMM are given by 
\begin{equation}
\mathbb{E}\left[\tau_{j}\right] = \frac{v_{jj} + \beta_j -1}{\beta_j - 1} \textrm{ and }  \textrm{Var}\left[\tau_{j}\right] = \frac{(v_{jj} + \beta_j -1)(v_{jj} + \beta_j - 2)}{(\beta_j - 1)(\beta_j - 2)} - \left(\frac{v_{jj} + \beta_j -1}{\beta_j - 1}\right)^2\nonumber
\end{equation}
for $\beta_j>2$ (the derivation of this result is provided in the Supplementary Material).

We therefore argue that a comparable prior specification  requires hyper-parameters $\{a_{0j}, b_{0j}\}_{j=1}^{K}$ and $\{\bm{v}_j \}_{j=1}^{K}$ be chosen in a way that satisfy $\mathbb{E}\left[\tau_{j}\right] = \mathbb{E}[\lambda_j + 1]$ and $\textrm{Var}\left[\tau_{j}\right] = \textrm{Var}\left[\lambda_j + 1\right]$, ensuring the dwell distribution in each state has the same prior mean and variance across models. The prior mean can be interpreted as a best a priori guess for the average dwell time in each state, and the variance reflects the confidence in this prior belief. In addition, since the negative binomial distribution is further parameterized by a dispersion parameter $\rho_j$, we center our prior belief at $\rho_j = 1$, which is the value that recovers geometric dwell durations (namely an \HMM) when $\lambda_j = \gamma_{jj}/(1-\gamma_{jj})$. Between state transition probabilities, i.e. the non-diagonal entries of the transition matrix, as well as the emission parameters, are  shared between the \HMM and  \HSMM, and thus we may place a prior specification on these parameters that is common across all models. 

%\jack{Stress that is is not that Bayes is better than Frequentist in general, it is Bayes is better than frequentist for this model.}

%\section{Simulation Studies} 
\section{A Comparison with Langrock and Zucchini (2011)}%: Prior Regularisation, Forecasting, Consistency and Complexity Penalization.}
\label{sec:simulation_studies}

This section presents several simulation studies. Firstly, we show that our Bayesian implementation provides similar point estimates as the methodology of \cite{langrock2011hidden}, serving as a ``sanity check''. We then proceed to illustrate the benefits adopting a Bayesian paradigm can bring to \HSMM modelling.

%Supplementary Material contains the \stan{} files, as well as \texttt{R} utilities, that were used to implement the Bayesian analysis for our experiments (see also \url{https://github.com/Beniamino92/BayesianApproxHSMM}). The probabilistic programming framework associated with \stan{} makes it easy for practitioners to consider further dwell/emission distributions to the ones considered here. Users need only to change the corresponding function in our \stan{} files. \jack{Do we not already say this in the intro}

%\jack{I wonder if we could trim this guy down, experimental details in the appendix, and just use it to focus on the prior regualrisation} \jack{Rather than Presenting Bayesian theory, describe the result of the experiments and the view them as a consequence of Bayesian theory: Present Experiment $\rightarrow$ Describe results $\rightarrow$ put in context of Bayesian theory}
%\vspace{-0.75cm}
%\textcolor{black}{\subsection{Inference and Prediction}}
\subsection{Parameter Estimation}
%\textcolor{black}{\subsection{Illustrative Example}}
%\textcolor{black}{\subsubsection{Prior Regularisation}}
\label{sec:ill_example}

\begin{figure}

\centering
\subfloat[]{
\centering
\includegraphics[scale = 0.5]{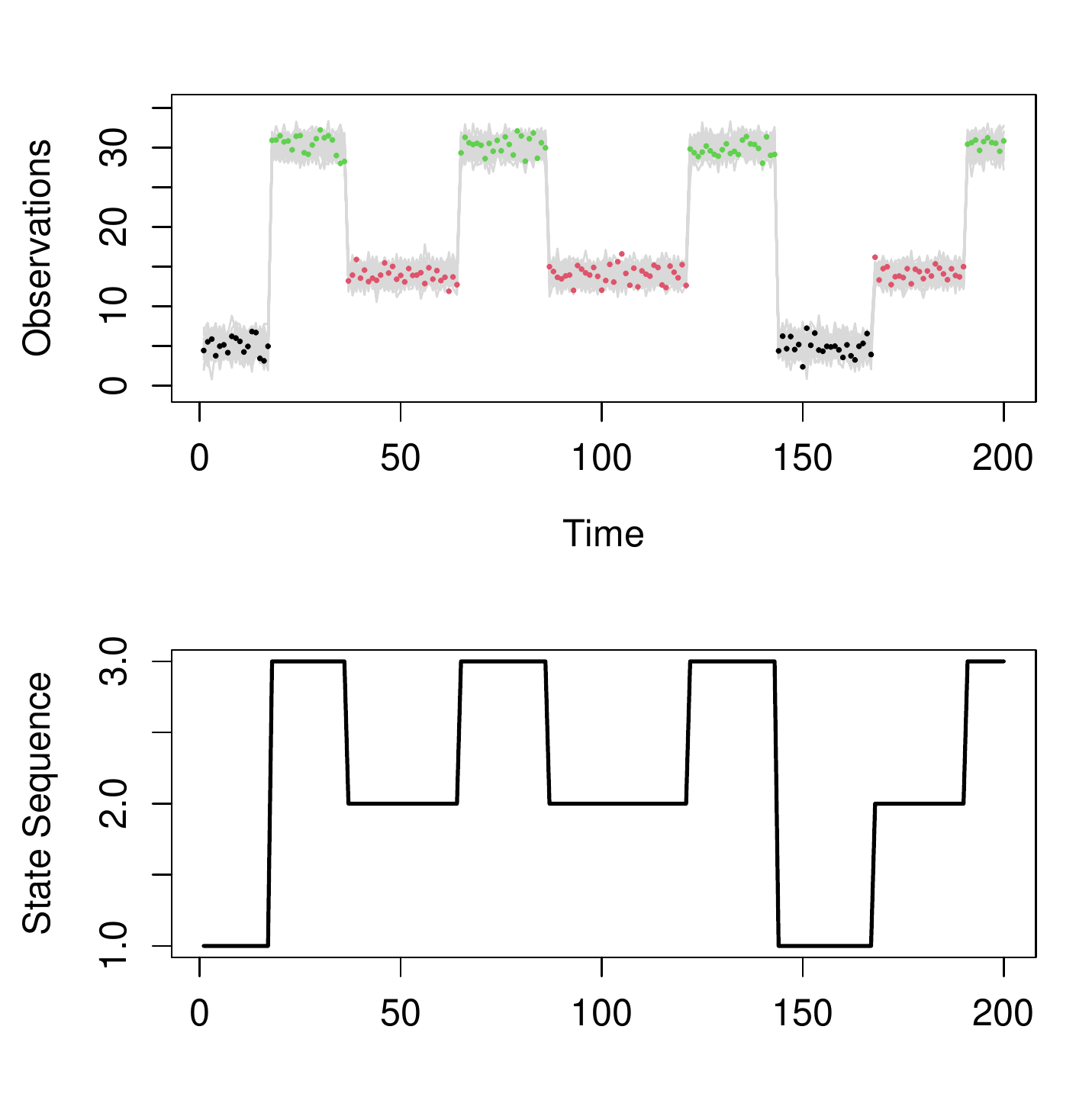}
}
%no space
\subfloat[]{
\centering
\includegraphics[scale = 0.5]{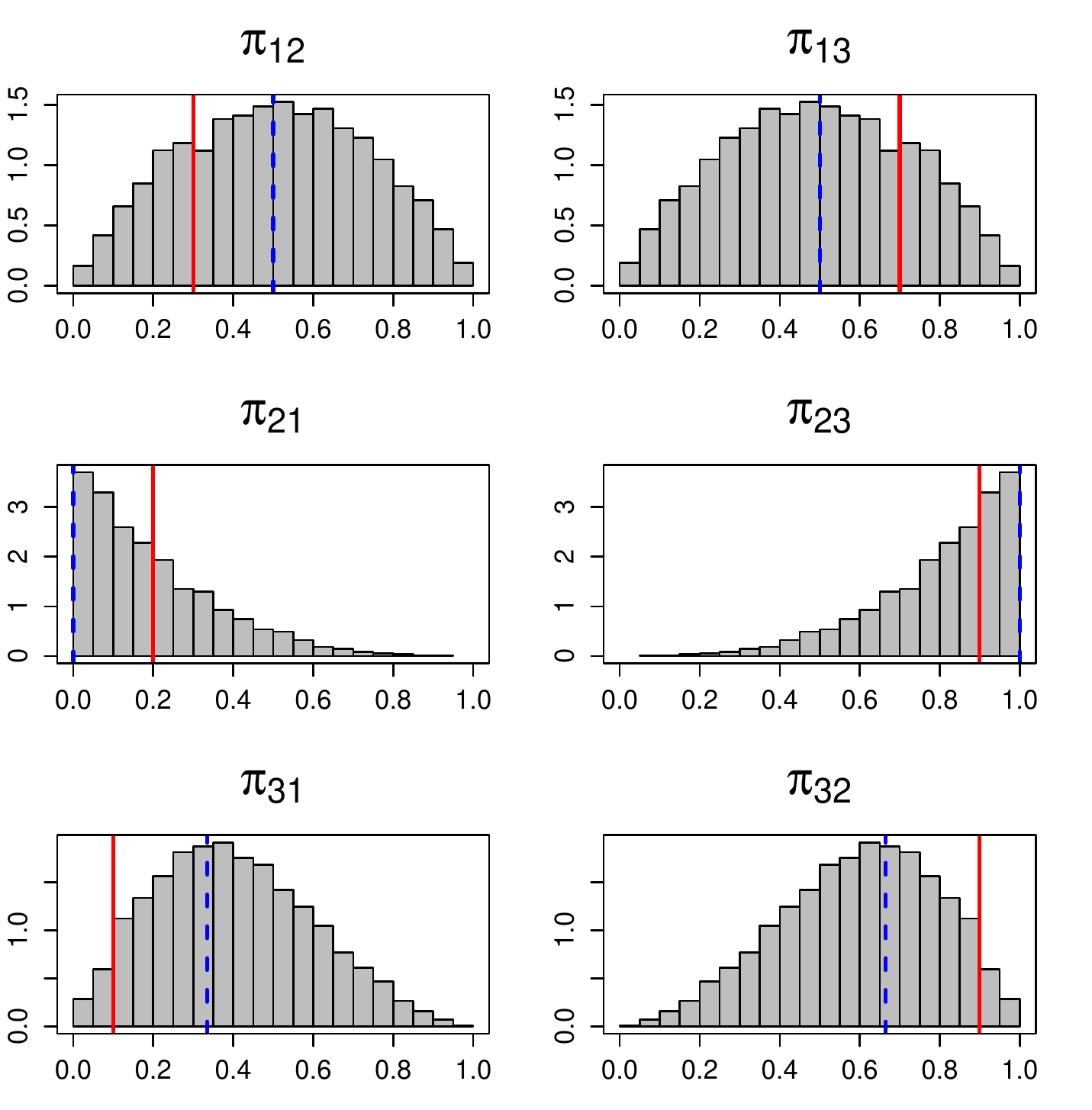}
}

\caption{(a, top) a realization (dots) of a three-state \HSMM with Gaussian emissions and Poisson durations, where different colors correspond to (true) different latent states. Grey lines represent 100 samples drawn from the estimated posterior predictive distribution.   
	(a, bottom) Most likely hidden state sequence estimated via the Viterbi algorithm; (b) estimated posterior distribution of the transition probabilities $\pi_{jk}$, where vertical solid red and blue dotted lines represent true values and EM estimates, respectively. }
\label{fig:trans_viterbi}
\end{figure}

For our first example, we simulated $T = 200$ data points from a three-state \HSMM (Eq. \ref{eq:HSMM}). Conditional on each state $j$, the observations are generated from a $\text{Normal}\left(\mu_j, \sigma_j^2\right)$, and the dwell durations are Poisson$(\lambda_j$) distributed. We consider relatively large values for $\lambda_j$ in order to evaluate the quality of the \HSMM approximation provided by Eq. \eqref{eq:bayes_model}. %with rates set to relatively high values. 
The full specification is provided in Table \ref{table:parms_and_res} and a realization of this model is shown in Figure \ref{fig:trans_viterbi} \textcolor{blue}{(a, top)}. The dwell approximation thresholds $\bm{a}$ are set equal to $(30, 30, 30)$ and we placed a Gamma$(0.01, 0.01)$ prior on the Poisson rates $\lambda_j$. The transition probabilities $\bm{\pi}_j$ are distributed as $\text{Dirichlet}(1, 1)$ and the priors for the Gaussian emissions are given as $\text{Normal}(0, 10^2)$ and $\text{Inverse-Gamma}(2, 0.5)$ for locations $\mu_j$ and scale $\sigma^{\,2}_j\,$, respectively. Overall, this prior specification is considered weakly informative \citep{gelman2013bayesian, gelman2017prior}. 

% Our proposed methodology required 2086 seconds to sample 6,000 iterations, 1,000 of which are discarded as burn-in, on an Intel\textsuperscript{\textregistered} Core\textsuperscript{TM} i7  2.2 GHz Processor 8 GB RAM. \jack{Is this relevant any more?}

Table \ref{table:parms_and_res} shows estimation results for our proposed  Bayesian methodology as well as the analogous  frequentist approach (\EM) of \citet{langrock2011hidden}, which will be referred to as LZ-2011.  %While it is clear that 
Figure \ref{fig:trans_viterbi}$\,$\textcolor{blue}{(a)} displays: (top) a graphical posterior predictive check consisting of the observations alongside 100 draws from the estimated posterior predictive  \citep{gelman2013bayesian};
%\jack{Can we put this in the appendix?}\jack{, the latter obtained by first drawing a sample path using a variant of the forward-backward procedure (see e.g. \citealt{hadj2020spectral}), and then, conditioned on the hidden state sequence, the predicted values are simulated from appropriate emission distributions. It is clear that no systematic discrepancies between observed and predicted data can be noticed;
(bottom) the most likely hidden state sequence, i.e. $ \argmax_{\bm{z}} p \, ( \, \bm{z} \, | \, \bm{y}, \,  \bm{\eta} \,    )$, which is estimated via the Viterbi algorithm (see e.g. \citealt{zucchini2017hidden}) using plug-in Bayes estimates of the model parameters; In order to assess the goodness of fit of the model, we also verified normality of the pseudo-residual (see Supplementary Material).

\begin{table}%[htbp]
	\centering 
	\resizebox{\columnwidth}{!}{%
		\begin{tabular}{lcccllcccllccc}
			\hline \\[-0.9em] 
			& True & LZ-2011    & Proposed                                                            &  &           & True & LZ-2011   & Proposed                                                         &  &        & True & LZ-2011   & Proposed                                                        \\ [.1em] \cmidrule{2-4} \cmidrule{7-9} \cmidrule{12-14}
			$\mu_1$    & 5    & 4.96  & \begin{tabular}[c]{@{}c@{}}4.95\\ \footnotesize(4.66–5.24)\end{tabular}    &  & $\sigma_3$  & 1    & 1.01 & \begin{tabular}[c]{@{}c@{}}1.08\\ \footnotesize(0.90–1.20)\end{tabular}  &  & $\pi_{13}$ & 0.70 & 0.50 & \begin{tabular}[c]{@{}c@{}}0.5\\ \footnotesize(0.13–0.87)\end{tabular}  \\
			$\mu_2$    & 14   & 14.02 & \begin{tabular}[c]{@{}c@{}}14.02\\ \footnotesize(13.67–14.37)\end{tabular} &  & $\lambda_1$ & 20   & 23.47 & \begin{tabular}[c]{@{}c@{}}23.36\\ \footnotesize(17.03–30.57)\end{tabular}  &  & $\pi_{21}$ & 0.20 & 0.00 & \begin{tabular}[c]{@{}c@{}}0.20\\ \footnotesize(0.01–0.53)\end{tabular} \\
			$\mu_3$    & 30   & 30.19 & \begin{tabular}[c]{@{}c@{}}30.18\\ \footnotesize(29.98–30.38)\end{tabular} &  & $\lambda_2$ & 30   & 27.22 & \begin{tabular}[c]{@{}c@{}}27.05 \\ \footnotesize(22.43–32.19)\end{tabular} &  & $\pi_{23}$ & 0.80 & 1.00 & \begin{tabular}[c]{@{}c@{}}0.80\\ \footnotesize(0.47–0.99)\end{tabular} \\
			$\sigma_1$ & 1    & 1.09  & \begin{tabular}[c]{@{}c@{}}1.15\\ \footnotesize(0.95–1.40)\end{tabular}     &  & $\lambda_3$ & 20   & 19.98 & \begin{tabular}[c]{@{}c@{}}20.00\\ \footnotesize(15.93–24.46)\end{tabular}  &  & $\pi_{31}$ & 0.10 & 0.33 & \begin{tabular}[c]{@{}c@{}}0.40\\ \footnotesize(0.10–0.76)\end{tabular} \\
			$\sigma_2$ & 2    & 1.90  & \begin{tabular}[c]{@{}c@{}}1.95\\ \footnotesize(1.73–2.22)\end{tabular}     &  & $\pi_{12}$    & 0.30 & 0.50 & \begin{tabular}[c]{@{}c@{}}0.50\\ \footnotesize(0.13–0.87)\end{tabular}  &  & $\pi_{32}$ & 0.90 & 0.67 & \begin{tabular}[c]{@{}c@{}}0.60\\ \footnotesize(0.24–0.90)\end{tabular} \\[.2em] \hline
		\end{tabular}%
	}
	\caption{Illustrative Example. True model parameterization and corresponding estimates obtained via the EM algorithm and our proposed Bayesian approach. For the latter, we also report  $95\%$ credible intervals estimated from the posterior sample. }
	\label{table:parms_and_res}
\end{table}

%\vspace{-0.75cm}
% \begin{figure}[htbp] 
% 	\centering
% 	\centerline{\includegraphics[scale = 0.28]{Plots/data_ill_ex.png}}
% 	\caption{Illustrative Example. A realization of a three-state \HSMM with Gaussian emissions and Poisson durations, where different colors correspond to different latent states.}
% 	\label{fig:data_ill_ex}
% \end{figure}

%\textcolor{black}{\subsubsection{Uncertainty Quantification and Prior Regularisation}}

In general, both methods satisfactorily retrieve the correct pre-fixed duration and emission parameters and the posterior predictive checks indicate that our posterior sampler is performing adequately. The implementation of \cite{langrock2011hidden} suffers from a lack of regularisation, for example in the estimation of $\pi_{21}$ as 0, and is not currently available with an automatic method to quantify parameter uncertainty. While augmenting the approach of \cite{langrock2011hidden} by adding regularisation penalties to parameters and producing confidence measures such as standard errors and bootstrap estimates is possible, such features are automatic to our Bayesian adaptation. Further, such an approach allows this uncertainty to be incorporated into methods for prediction and model selection making the Bayesian paradigm appealing for \HSMM modelling.

\color{black}
%\jack{can you make a little intro? This section illustrates our improvements over \citet{langrock2011hidden} in terms of forecasting and model selection. }

\subsection{Forecasting}

%\jack{One reason I think this might be is that we have many parameters and the possibility that many parameters are influenced by few observations}

\begin{figure}[t]%[htbp] 
	\centering
	%\centerline{\includegraphics[scale = 0.42]{Plots/forecast_boxplots.pdf}}
	\centerline{\includegraphics[scale = 0.5]{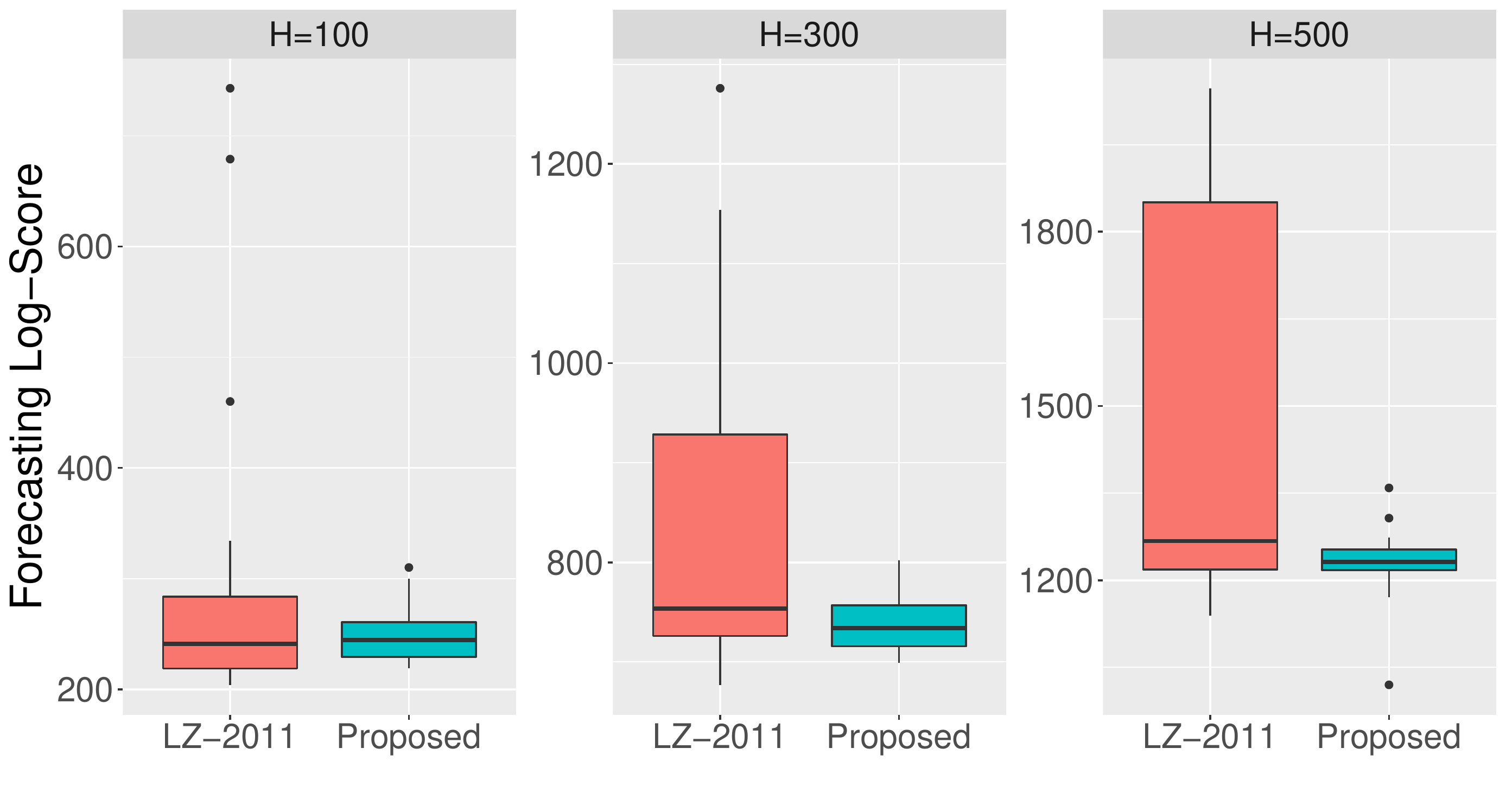}}
	\caption{Boxplots of log-scores for LZ-2011 (via \EM) and our Bayesian methodology, with three different forecast horizons $H = 100, 300, 500$. %It is evident that our Bayesian approach is superior in terms of forecasting performance.
	}
	\label{fig:boxplots_forecast}
\end{figure}

A key feature of \HSMMs is their ability to be able to capture and forecast when and for how long the model will be in a given state. We compare the forecasting properties of the method presented by \citet{langrock2011hidden} and our proposed Bayesian approach. We simulated 20 `un-seen' time series, $\tilde{\bm{y}} = (\tilde{y}_{\,1}, \, \ldots, \tilde{y}_H)$, where $\tilde{y}_{\,h} = y_{\, T +h}, \hspace{0.1cm} h = 1, \dots, H$ and $H = 100, 300, 500$ denotes the forecast horizon, from the model as in Table \ref{table:parms_and_res}. We used the logarithmic score (log-score)  to  measure  predictive performances. %, while acknowledging that other scoring rules may be equally valid. 
Let  $\hat{\bm{\eta}}$ be the frequentist  (\MLE/\EM)  parameter estimate and define the log-score \begin{equation*}
    L_{\,\text{freq}}(\tilde{\bm{y}}) = \sum_{h=1}^{H} -\log p \, (\tilde{y}_h \, | \, \hat{\bm{\eta}} ),
\end{equation*} where  $p \,(\tilde{y}_h \, | \, \hat{\bm{\eta}})$ denotes  the forecast density function (see Supplementary Material for an explicit expression). Our Bayesian framework does not assume a point estimate $\hat{\bm{\eta}}$ but considers instead a posterior distribution $p \, (\bm{\eta} \, | \, \bm{y})$, %. Rather than making predictions using a plugin estimate, we integrate out the model parameters and  evaluate the predictive density of a future observation $\tilde{y}_{\,h}$. 
which is integrated over to produce a predictive density.
%This is achieved by approximating the following predictive distribution
%\begin{equation*}
%    p \, ( \tilde{y}_h \, | \, \bm{y}) = \int p \, (\tilde{y}_{h} \, | \, \bm{\eta} ) \, p \, (\bm{\eta} \, | \, \bm{y}) \, d\bm{\eta} 
%    \approx \frac{1}{M}\sum_{i = 1}^M p \, (\tilde{y}_h \, | \, \bm{\eta}^{(i)} ),
%\end{equation*} where $\bm{\eta}^{(i)}$ represent a draw from $p \, (\bm{\eta} \, | \, \bm{y})$ and $M$ is total number of \MCMC iterations. As a result, the Bayesian predictive log-score becomes
%\begin{equation*}
   % L_{\,\text{Bayes}}(\tilde{\bm{y}}) = \sum_{h=1}^{H} \log\left( \frac{1}{M}\sum_{i = 1}^M p \, (\tilde{y}_{h} \, | \, \bm{\eta} ) \, p \, (\bm{\eta}^{(i)} \, | \, \bm{y})\right), \quad \left\lbrace\bm{\eta}^{(i)}\right\rbrace_{i=1}^M \sim \pi \, (\bm{\eta} \, | \, \bm{y}).
%   L_{\,\text{Bayes}}(\tilde{\bm{y}}) = \sum_{h=1}^{H} \log\left( \frac{1}{M}\sum_{i = 1}^M p \, (\tilde{y}_{h} \, | \, \bm{\eta} )\right), \quad \left\lbrace\bm{\eta}^{(i)}\right\rbrace_{i=1}^M \sim \pi \, (\bm{\eta} \, | \, \bm{y}).
%\end{equation*}
%
Given $M$ \MCMC samples drawn from  the posterior, $\left\lbrace\bm{\eta}^{(i)}\right\rbrace_{i=1}^M \sim \pi \, (\bm{\eta} \, | \, \bm{y})$, the log-score of the predictive density can be approximated as
\begin{align*}
    L_{\,\text{Bayes}}(\tilde{\bm{y}}) = \sum_{h=1}^{H} -\log p \, ( \tilde{y}_h \, | \, \bm{y}) &= \sum_{h=1}^{H} -\log\int p \, (\tilde{y}_{h} \, | \, \bm{\eta} ) \, p \, (\bm{\eta} \, | \, \bm{y}) \, d\bm{\eta} \\
    &\approx \sum_{h=1}^{H} -\log\left( \frac{1}{M}\sum_{i = 1}^M p \, (\tilde{y}_h \, | \, \bm{\eta}^{(i)} )\right).
\end{align*}
Figure \ref{fig:boxplots_forecast} presents box-plots of log-scores for LZ-2011 and our proposed Bayesian approach. It is clear that our Bayesian methodology typically produces a much lower predictive log-score than the frequentist procedure. The approach by \citet{langrock2011hidden}  which uses plug-in estimates for parameters, is known to `under-estimate' the true predictive variance  %(\jack{reference}) 
thus yielding large values of the log-score \citep{jewson2018principles}. On the other hand, our Bayesian paradigm integrates over the parameters and hence is more accurately able to capture the true forecast distribution. As a result, it produces significantly smaller log-score estimates. %In fact, \citet{aitchison1975goodness} showed that, under repeat sampling, the Bayesian predictive is closer, in terms of Kullback-Leibler divergence, to the true underlying distribution than its frequentist plug-in counterpart.

% It becomes clear from this figure that choosing a Bayesian paradigm produces generally lower predictive scores than a frequentist procedure. the advantage of the Bayesian paradigm's full quantification of uncertainty in producing generally lower predictive scores than the frequentist procedure. By considering plug-in estimates for parameters, the frequentist approach can be seen to `under-estimate' the true predictive variance (\jack{reference}), something which is known to be heavily penalised by the log-score \citep{jewson2018principles}. This is expected and explains the extremely high values for the log-score sometimes obtained under a frequentist approach. On the other hand, our Bayesian paradigm integrates over the parameters and is more accurately able to capture the true forecast distribution and as a result, it produces smaller log-score estimates

%  Given the flexibility of the \HSMM formulation in Section \ref{sec:model} to extend the dwell distribution beyond the geometric distribution implied by the \HMM, an important consideration is how to select between such approaches in a data-driven manner. 

%\newpage
\textcolor{black}{\subsection{Dwell Distribution Selection}}

An important consideration is whether to formulate an $\HMM$ or to extend the dwell distribution beyond a geometric one (i.e., an \HSMM). Ideally, the data should be used to drive such a decision.  In this section, we compare the frequentist methods for doing so, namely Akaike's information criterion (\AIC, \citealt{akaike1973information}) and Bayesian information criterion (\BIC, \citealt{schwarz1978estimating}), with their Bayesian counterpart, namely the marginal likelihood. We choose not to consider other Bayesian inspired information criteria \citep[e.g.][]{spiegelhalter2002bayesian, watanabe2010asymptotic, gelman2014understanding} as our goal here is to compare standard frequentist methods used previously in the literature to conduct model selection for \HMMs and \HSMMs \citep[e.g.][]{langrock2011hidden, huang2018hidden} with the canonical Bayesian analog.  Although the performance of Bayesian model selection can be sensitive to the specification of the prior, we gave specific consideration to specifying this with model selection in mind in Section \ref{sec:comparable_dwell}.

\textcolor{black}{\subsubsection{Consistency for Nested Models}}

A special feature of the negative binomial dwell distribution is that the geometric dwell distribution associated with \HMMs is nested within it. Taking $\rho = 1$ for the negative binomial exactly corresponds to the geometric distribution.  An important consideration when selecting between nested models is complexity penalization. For the same data set, the more complicated of two nested models will always achieve a higher in-sample likelihood score than the simpler model. Therefore, in order to achieve consistent model selection among nested models, the extra parameters of the more complex models must be penalized. %, as such a model, is guaranteed to achieve higher values for the maximised log-likelihoods when evaluated on the same data. %As a result, simpler models can only be selected if the complexity of the models is somehow penalized.  
In this scenario, the \AIC \,:= -2\,$\mathscr{L}\,( \bm{y} \, | \,  \bm{\eta} ) + 2 p$ where $p$ denotes the number of parameters included in the model, is known not to provide consistent model selection when the data is generated from the simpler model \citep[see e.g.][]{fujikoshi1985selection}. %This metric penalizes complexity through the term $2p$, which does not depend on the number of data points $T$. Thus, as $T$ tends to infinity there is still a chance that an increased likelihood can overwhelm the complexity penalty, yielding a non-zero probability that \AIC will incorrectly select the more complicated model.
%\AIC cannot guarantee to select the simpler data generating model with probability 1. 
On the other hand, performing model selection using the marginal likelihood can be shown to be consistent (see e.g. \citealt{o2004kendall}), provided some weak conditions on the prior are satisfied. Therefore, when following a Bayesian paradigm, the correct data generating model is selected with probability one as $T$ tends to infinity.  Here we show that under the approximate \HSMM likelihood model, Bayesian model selection appears to maintain its desirable properties.

We simulated 20 time series from a two-state \HMM with Gaussian emission parameters $\bm{\mu} = (1, 4)$ and $\bm{\sigma}^2 = (1, 1.5)$, and diagonal entries of the transition matrix set to $(\gamma_{11}, \gamma_{22}) = (0.7, 0.6)$. To model this data we considered the \HMM and a \HSMM with negative binomial durations \color{black}. For the \HSMM approximation, we considered $\bm{a} = (3,3)$, $(5, 5)$ and $(10, 10)$ in order to investigate how the dwell approximation affects the model selection performance. We use prior distributions that are comparable as explained in Section \ref{sec:comparable_dwell}, the exact prior specifications are presented in the Supplementary Material. Figure \ref{Fig:BFs_AIC_BIC} (top) displays box-plots  of the difference between the model selection criteria (namely marginal likelihood and \AIC) achieved by the \HMM and the \HSMM, for increasing sample size $T = 500, 5000, 10000$ and values for $\bm{a}$. We negate the \AIC such that maximising both criteria is desirable. Thus, positive values for the difference correspond to correctly selecting the simpler data generating model, i.e. the \HMM. As the sample size $T$ increases, the marginal likelihood appears to converge to a positive value, and the variance across repeats decreases, indicating consistent selection of the correct model. On the other hand, even for large $T$ there are still occasions when the \AIC strongly favours the incorrect, more complicated model. Further, such performance appears consistent across values of $\bm{a}$.%, which is a symptom of the lack of consistency exhibited by \AIC.  %Consistently selection of simple models has importance both inferentially, in terms of understanding the dwell distribution of the data and also computationally, the \HMM is computationally cheaper than simple models for computational and inferential convenience.

\begin{figure}[htbp] 
	\centering
	\centerline{
	\includegraphics[width = 0.8\columnwidth]{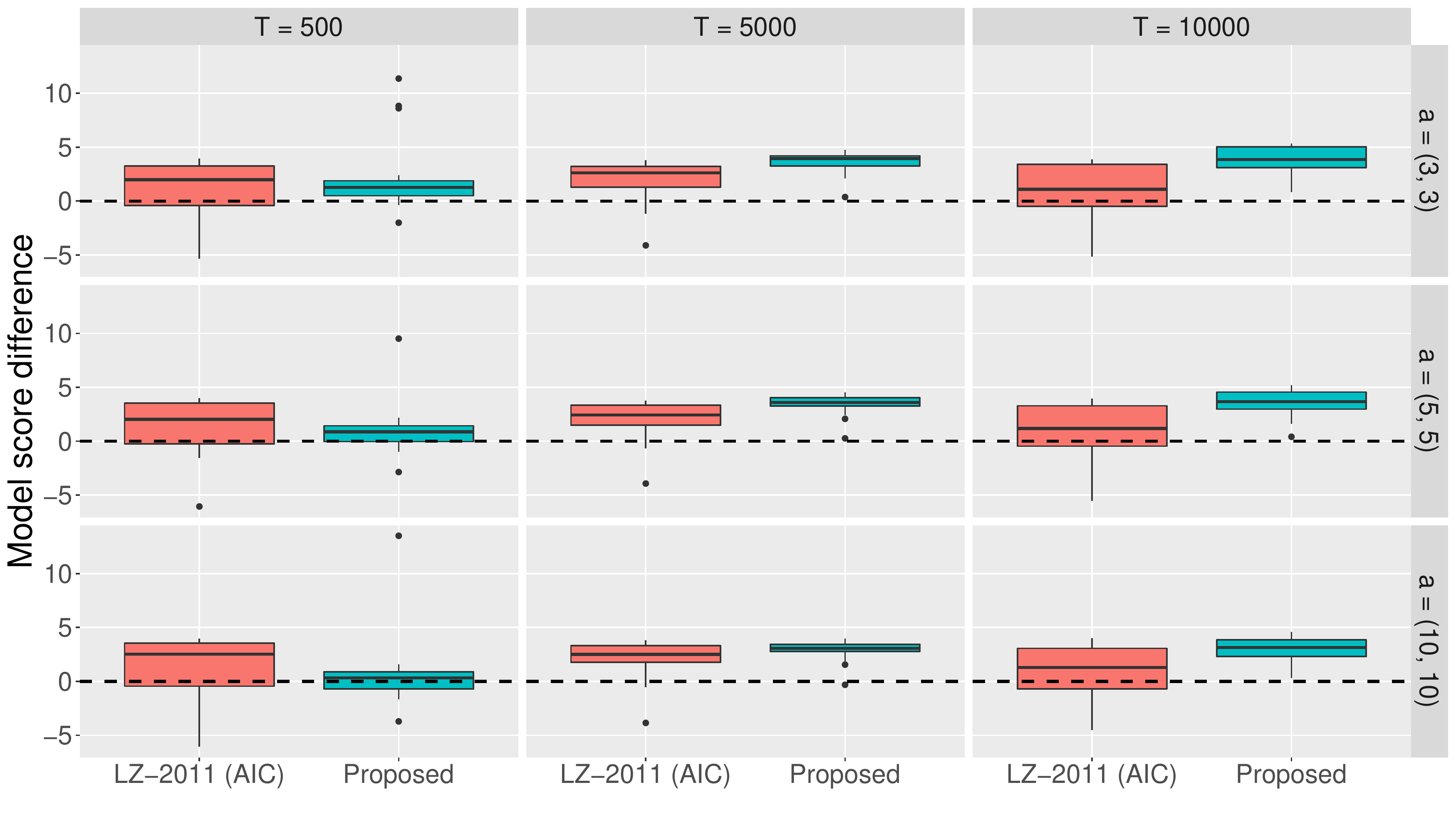}
	}
	\centerline{
	\includegraphics[width = 0.8\columnwidth]{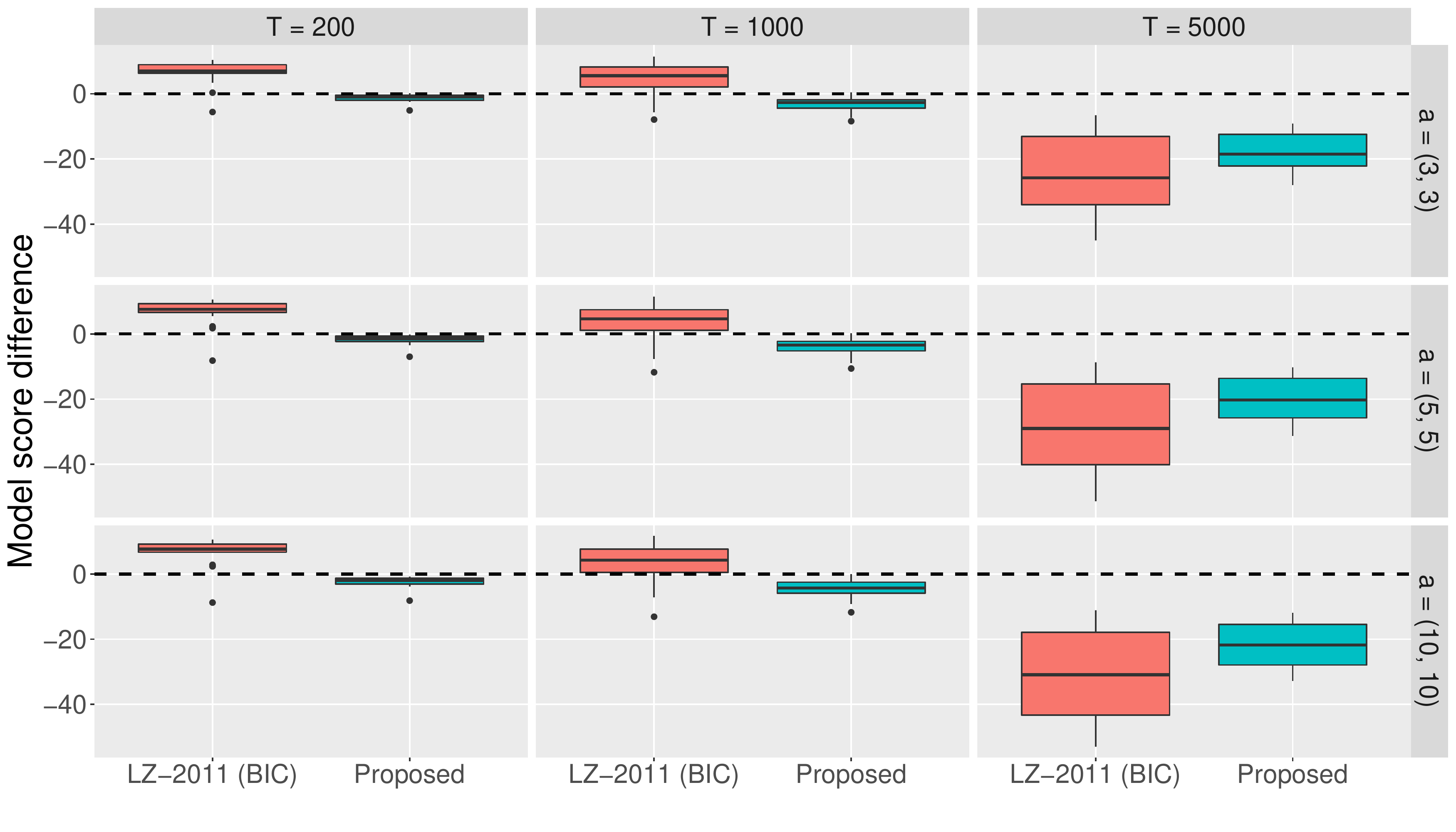}
	}
	\caption{ \AIC, \BIC and the marginal likelihood for nested models. (Top) model score differences between a negative binomial duration \HSMM with $\bm{a} = (3, 3), (5, 5)$ and $(10, 10)$ and an \HMM when the data is generated from the \HMM. Positive values of the model score difference correspond to correctly selecting the simpler model. (Bottom) model score differences between a negative binomial duration \HSMM  approximated by a threshold $\bm{a} = (3, 3), (5, 5)$ and $(10, 10)$ and an \HMM when the data is generated from the \HSMM. Negative values of the model score difference correspond to correctly selecting the more complicated model. 
	%Note that here we interpret the methods as being quantitatively comparable tools for model selection. That is to say that, the difference between the \AIC/\BIC and the log-marginal
    %likelihood values of the models are comparably used for model selection decisions,
    %with being greater than or less than 0 corresponds to the selection decision. 
    Note that here we interpret the difference between the \AIC/\BIC and the log-marginal likelihood values of two models as quantitatively comparable for model selection decisions, with being greater than or less than 0 corresponding to the selection decision.
    }
	\label{Fig:BFs_AIC_BIC}
\end{figure}

\textcolor{black}{\subsubsection{Complexity Penalization}}

Unlike the \AIC, the $\BIC \,:= -2\,\mathscr{L}\,( \bm{y} \, | \,  \bm{\eta} ) + p \log T$  penalizes complexity in a manner that depends on the sample size $T$.  This is termed `Bayesian' because it corresponds to the Laplace approximation of the marginal likelihood of the data \citep{konishi2008information}, often interpreted as considering a uniform prior for the model parameters \citep{bhat2010derivation, sodhi2010conservation}. Though the uniform distribution may be viewed as naturally uninformative, it is well known that using the marginal likelihood assuming an uninformative prior specification can lead to the selection of the simplest model independently of the data \citep[see e.g.][]{lindley1957statistical, jeffreys1998theory, jennison1997bayesian}. As a result, while  \BIC can provide consistent selection of nested models, it can punish extra complexity in an excessive manner. 

To investigate how the approximate \HSMM likelihood model affects this model selection behaviour,  we consider data generated from an \HSMM with the same formulation as above except that in this scenario the dwell distribution is a negative binomial parameterized by state-specific parameters $\bm{\lambda} = (3.33, 2.50)$ and $\bm{\rho} = (2, 0.5)$. Note that the data generating \HSMM has two more parameters than the \HMM. For the \HSMM approximation, we consider $\bm{a} = (3,3)$, $(5, 5)$ and $(10, 10)$, where the largest of these provides negligible truncation of the right tail of the dwell distribution given the data generating parameters.  Figure \ref{Fig:BFs_AIC_BIC} (bottom) shows box-plots  of the difference between the model scores (marginal-likelihood and \BIC) across 20 simulated time series when fitting the \HMM and \HSMM, for increasing sample size $T = 200, 1000, 5000$ and values for $\bm{a}$. We negate the \BIC so that the preferred model maximises both criteria. Unlike the experiments described above, the data is now from the less parsimonious \HSMM approach and therefore negative values for the difference in score correspond to correctly selecting the more complicated model. For small sample sizes, e.g. $T = 200, 1000$, the complexity penalty of the \BIC appears to be too large, so that in almost all of the 20 repeat experiments the simple model is incorrectly favored over the correct data generating model, i.e. the \HSMM. On the other hand, the marginal likelihood is able to correctly select the more complicated model across almost all simulations and sample sizes. Although for smaller $\bm{a}$ the \HSMM approximation is `closer' to a \HMM, we still see that the model selection performance is consistent across the different values of $\bm{a}$.

%The supplementary material contains results of an additional experiment considering data from a negative binomial dwell distribution as well as full details of the prior specification. 

%Naturally, the results of this section are dependent upon the prior specifications for the marginal likelihood. What we aim to demonstrate here are some of the drawbacks of using \AIC and \BIC for model selection (as in \citealt{langrock2011hidden}),  and that, provided careful consideration is given to the prior specification, both large sample consistency and small sample efficiency can be instead achieved when following our Bayesian paradigm. 

\section{Approximation Accuracy and Computational Time}
%\subsection{Approximation Accuracy and Computational Time}
\label{sec:approx_accuracy}

The previous section motivated why the Bayesian paradigm can improve statistical inferences for \HSMMs. Next, we investigate the computational feasibility of such an approach and the trade-off between computational efficiency and statistical accuracy achieved by our Bayesian approximate \HSMM implementation. In particular, we compare our Bayesian approximate \HSMM method for different values of the threshold $\bm{a}$ with a Bayesian implementation of the exact \HSMM, while also illustrating the computational savings made by our sparse matrix implementation. For the exact \HSMM, the full-forward recursion is used to evaluate the likelihood (see e.g. \citealt{guedon2003estimating} or \citealt{economou2014mcmc}). In order to provide a fair comparison, we coded the forward recursion outlined in \citet{economou2014mcmc} in \stan{} also. We then compare the computational resources required to sample from the approximate and exact \HSMM posteriors with the accuracy of the posterior mean parameter estimates with respect to their data generating values. 

% We further illustrate the computational savings made by using the sparse matrix implementation.

We generate $T = 5000$ observations from two different \HSMMs with Poisson durations (both with $K = 5$ states and the same Gaussian emission distributions). For the two different datasets, we consider the following dwell parameters: (i) \textit{short dwells}, i.e. $\bm{\lambda} = (2, 5, 8, 1, 4)$, where the average time spent in each state is fairly small and (ii) \textit{one long dwell}, i.e. $\bm{\lambda} = (2, 5, 25, 1, 4)$,  where four states have short average dwell time and one where the average dwell time is much longer.
%\begin{enumerate}
%    \item Short dwells: $\lambda = (2, 5, 8, 1, 4)$
%    \item One long dwell: $\lambda = (2, 5, 25, 1, 4)$
%\end{enumerate}
%
%In regime 1, the average time spent in each state is fairly small (all $< 8$), while regime 2 has 4 states where the average dwell time is short and one where the average dwell time is much longer. 
%
 We also consider two approximation thresholds: $\bm{a}_1 = (10, 10, 10, 10, 10)$, namely a fixed approximation threshold for all five states, and $\bm{a}_2 = (10, 10, 30, 10, 10)$, a `hybrid' model where four of the states have short dwell thresholds and one has a longer threshold.
%\begin{itemize}
%    \item Constant threshold: $a_1 = (10, 10, 10, 10, 10)$
%    \item Hybrid threshold: $a_2 = (10, 10, 30, 10, 10)$
%\end{itemize}
%The first \HSMM approximation fixes the approximation threshold at 10 for all states while the second has it at 10 for 4 of the states and having one state with a higher threshold \jack{hybrid model}. 
The emission parameters were set to $\bm{\mu} = (1, 2, 3.5, 6, 10)$ and $\bm{\sigma}^2 = (1^2, 0.5^2, 0.75^2, 1.5^2, 2.5^2)$, and we specify priors $\mu_j \sim \mathcal{N}(0, 10^2)$, $\sigma^2_j \sim \mathcal{IG}(2, 0.5)$, $\lambda_j\sim\mathcal{G}(0.01, 0.01)$ and $\gamma_j \sim \mathcal{D}(1, \ldots, 1)$ for $j = 1,\ldots, 5$.

The results are presented in Table \ref{Tab:Computation_vs_Accuracy}. Across both datasets and approximation thresholds, the sparse implementation takes less than half the time of the non-sparse implementation, with the saving greater when the dwell thresholds are larger (and the matrix $\bm{\Phi}$, Eq. \eqref{eq:diag_mat}, is sparser).  Furthermore, the \HSMM approximations are considerably faster than the full \HSMM implementation. For the \textit{short dwell} dataset the full \HSMM takes close to 3.5 days while the sparse implementations of the \HSMM approximation both require less than 2 hours. Similarly, for the \textit{one long dwell} dataset, the full \HSMM takes over 4 days to run while again the sparse \HSMM approximations require around 2 hours. The quoted Effective Sample Size (\ESS, e.g. \citealt{gelman2013bayesian}) values are calculated using the \textit{LaplaceDemons} package in R and are averaged across parameters. These show that the \ESS of all the generated samples is close to 1000 and thus the time comparisons are indeed fair. Further, we expect the difference to become starker as the number of observations $T$ increases. While, the approximate \HSMM scales linearly in $T$ and quadratically in $\sum_{j=1}^Ka_j$, the full \HSMM in the worst case is quadratic in $T$ \citep{langrock2011hidden}.

Lastly, we see that the savings in computation time come at very little cost in statistical accuracy. We measure the statistical accuracy of the vector-valued parameter  $\hat{\bm{\theta}}$ to estimate $\bm{\theta}^{\ast}$ using its mean squared error (\MSE 
=   $\sum_{j=1}^K(\hat{\theta}_j - \theta^{\ast}_j)^2$). All methods achieve almost identically \MSE values for the emission parameters $
\bm{\mu}$ and $\bm{\sigma}^2$. For the \textit{short dwell data}, the $\bm{a}_1$ approximation has slightly higher \MSE for $\bm{\lambda}$ while the $\bm{a}_2$ approximation performed comparably to the \HSMM. Clearly, increasing the approximation threshold improves statistical accuracy. On the other hand, the \textit{one long dwell} shows that if the dwell threshold is set too low, as is the case with $\bm{a}_1$, large errors in the dwell estimation can be made. %, thus care should be taken in practice
However, in this example the higher dwell approximation $\bm{a}_2$ once again performs comparably with the full \HSMM, whilst requiring only $2\%$ of the computational time.

\begin{table}[ht]
\centering
\begin{tabular}{lccccc}
  \hspace{-0.2cm}\\
  \hline
  \textit{Short dwells} & Time (hours) & \ESS & \multicolumn{3}{c}{\MSE}\\
   & & & $\mu$ & $\sigma^2$ & $\lambda$ \\ 
  \hline
  Approx: $\bm{a_1}$ & 2.62 & 986.50 & 3.72 $\times 10^{-2}$ & 2.88 $\times 10^{-3}$  & 0.25 \\ 
  Approx ({\footnotesize{SPARSE}}): $\bm{a_1}$ & 1.30 & 975.20 & 3.84 $\times 10^{-2}$ & 3.06 $\times 10^{-3}$ & 0.26 \\ 
  Approx: $\bm{a_2}$ & 3.94 & 961.76 & 3.84 $\times 10^{-2}$ & 3.05 $\times 10^{-3}$ & 0.17 \\ 
  Approx:  ({\footnotesize{SPARSE}}): $
  \bm{a_2}$ & 1.78 & 978.82 & 3.96 $\times 10^{-2}$ & 3.01 $\times 10^{-3}$ & 0.18 \\ 
  Exact & 81.15 & 933.28 & 4.02 $\times 10^{-2}$ & 3.24 $\times 10^{-3}$ & 0.19 \\
\hline
%\\
%\end{tabular}
%\end{table}
%\begin{table}[ht]
%\centering
%\begin{tabular}{lccccc}
\hline
  \textit{One long dwell} & & & & &\\
  %textit{One long dwell} & Time (hours) & ESS & \multicolumn{3}{c}{\MSE}\\
  %& & & $\mu$ & $\sigma^2$ & $\lambda$ \\ 
  \hline
  Approx: $\bm{a_1}$ & 3.33 & 984.51 & 1.76 $\times 10^{-2}$ & 4.68 $\times 10^{-2}$ & 128.50 \\ 
  Approx:  ({\footnotesize{SPARSE}}): $\bm{a_1}$ & 1.78 & 981.90 & 1.73 $\times 10^{-2}$ & 4.84 $\times 10^{-2}$ & 128.51 \\ 
  Approx: $
  \bm{a_2}$ & 5.08 & 993.89 & 1.50 $\times 10^{-2}$ & 4.84 $\times 10^{-2}$ & 1.25 \\ 
  Approx:  ({\footnotesize{SPARSE}}): $\bm{a_2}$ & 2.21 & 983.47 & 1.51 $\times 10^{-2}$ & 4.82 $\times 10^{-2}$ & 1.25 \\ 
  Exact & 101.35 & 980.59 & 1.65 $\times 10^{-2}$ & 4.66 $\times 10^{-2}$ & 1.12 \\ 
\hline
\end{tabular}
\caption{
\textcolor{black}{Computational time (hours), effective sample size (\ESS) and  mean squared error (\MSE) of posterior mean parameters. The results are reported using the approximate \HSMM for different dwell approximations $\bm{a}$ (with the corresponding sparse implementation), and the exact \HSMM implementation.}
}
\label{Tab:Computation_vs_Accuracy}
\end{table}

%\color{black}

\subsection{Setting the Dwell Threshold}{\label{sub:setting_dwell}}

%\jack{$\bm{\eta} = \big\{ \, (  \bm{\pi}_j, \, \bm{\lambda}_j, \, \bm{\theta}_j ) \, \big\}_{j=1}^{\,K}$}

The results of Section \ref{sec:approx_accuracy} indicate that while vast computational savings are possible using the approximate \HSMM likelihood, care must be taken not to set the dwell approximation threshold $\bm{a}$ too low. We propose initialising $\bm{a}$ based on the prior distribution for the dwell times $d_j$, $\pi(d_j) = \int \pi(d_j; \lambda)\pi(\lambda)d\lambda$. Noting that any dwell time $d_j < \bm{a}_j$ is not approximated, we recommend initialising $\tilde{\bm{a}}$ such that $d_j \leq \tilde{\bm{a}}_j$ with high probability for all $j = 1,\ldots, K$.

Such an initialisation however does not guarantee the accuracy of the \HSMM modelling, particularly in the absence of informative prior beliefs. %Inspired by %\jack{MCMC/ABC/Calibration diagnostics (\textbf{references})}
We therefore, propose a diagnostic method to check that $\tilde{\bm{a}}$ is not too small. 
\begin{enumerate}[noitemsep]
	\item Initialise $\tilde{\bm{a}}$ and conduct inference on the observed data. Record posterior mean parameter estimates $\hat{\bm{\eta}}_{obs}(\tilde{\bm{a}})$%for emission distribution parameters $\hat{\theta}_{obs}(\tilde{\bm{a}})$ and dwell distribution parameters $\hat{\lambda}_{obs}(\tilde{\bm{a}})$
	\item Generate data $\tilde{y}_{gen}$ from an exact \HSMM with generating parameters $\hat{\bm{\eta}}_{obs}(\tilde{\bm{a}})$. Note that generation from an exact \HSMM is easier than inference on its parameters
	\item Continuing with $\tilde{\bm{a}}$, conduct inference on the generated data and record posterior mean parameter estimates $\hat{\bm{\eta}}_{gen}(\tilde{\bm{a}})$
	\item Compare dwell distribution parameters $\hat{\lambda}_{obs}(\tilde{\bm{a}})$ and $\hat{\lambda}_{gen}(\tilde{\bm{a}})$
\end{enumerate}
The estimates $\hat{\lambda}_{obs}(\tilde{\bm{a}})$ provide the best guess estimate of the parameters of the \HSMM underlying the data for fixed $\tilde{\bm{a}}$. Generating from this exact \HSMM given by these estimates allows us to verify the accuracy of the proposed model. If the estimates are not accurate then little confidence can be had that $\hat{\lambda}_{obs}(\tilde{\bm{a}})$ accurately represents the dwell distribution of the underlying \HSMM. If $\hat{\lambda}_{gen}(\tilde{\bm{a}})_j$ is not considered a satisfactory estimate of $\hat{\lambda}_{obs}(\tilde{\bm{a}})_j$, then $\tilde{\bm{a}}_j$ must be increased. Conveniently, this can be done for each state $j$ independently. Further, if $\hat{\lambda}_{gen}(\tilde{\bm{a}})_j$ is considered accurate enough, then there is also the possibility to decrease $\tilde{\bm{a}}_j$ based on the inferred dwell distribution. Although the above procedure requires the fitting of the model several times, we believe the computational savings of our model when compared with the exact \HSMM inference demonstrated in Table \ref{Tab:Computation_vs_Accuracy} render this worthwhile. %One could alternatively consider $a$ as simply a parameter of the dwell distribution and consider model selection for $a$
 This procedure is implemented to set the dwell-approximation threshold for the physical activity time series analysed in the next section.  
\color{black}

%The motivation for such a check is as follows: $\hat{\theta}_{real}(\tilde{a})$ provides the best guess estimate of the parameters of the \HSMM underlying the data (approximated according to $\tilde{a}$. We then generated data from this best fitted \HSMM and see if we can recover the data generating parameters using the same threshold $\tilde{a}$. If this cannot be done, then we have little confidence that $\hat{\theta}_{real}(\tilde{a})$ accurately estimates the underlying \HSMM.

% Need to define MSE, EES or something like that. 
% Need a sentence on the method of Economou et al.

% A sentence why you have threshold central (30), related to the hybrid model}

\section{Telemetric Activity Data}
\label{sec:application}

% presented in Figure \ref{fig:physical_activity} 

In this section, we return to the physical activity (\PA) time series that \citet{huang2018hidden} analysed using a frequentist \HMM. 
\color{black}
%In this section, we analyze the physical activity (\PA) time series that was investigated using a frequentist \HMM by  \citet{huang2018hidden}. 
We seek to conduct a similar study but within a Bayesian framework and consider the extra flexibility afforded by our proposed methodology to investigate departures from the \HMM. \textcolor{black}{Further, in Section \ref{sec:harmonic_emissions} we consider the inclusion of spectral information within the \HMM and \HSMM emission densities.}%, and hence study how this affects the \HMM and \HSMM models investigated here.}

We consider three-state  $\HSMMs$  with Poisson$\,(\lambda_j)$ and Neg-Binomial$\,(\lambda_j, \rho_j)$ dwell durations, shifted to have strictly non-negative support and approximated  via thresholds $\bm{a}_{P} = (160, 40, 25)$ and $\bm{a}_{NB} = (250, 50, 50)$ respectively. These are fitted to the square root of the \PA time series shown in Figure \ref{fig:physical_activity}, wherein we assume that transformed observations are generated from Normal$(\,\mu_j, \sigma_j^{\,2})$ distributions, as in \citet{huang2018hidden}. We specified $K = 3$ states, in agreement with findings of \citet{migueles2017accelerometer} and \citet{huang2018hidden}, where they collected results from more than forty experiments on \PA time series. In their studies, for each individual the lowest level of activity corresponds to the sleeping period, which usually happens during the night, while the other two phases are mostly associated with movements happening in the daytime. Henceforth, these different telemetric activities are represented as  inactive (\IA), moderately active (\MA) and highly active (\HA) states. The setting of $\bm{a}$ followed the iterative process outlined in Section \ref{sub:setting_dwell}, initialising $\tilde{a}_j$ giving prior probability of 0.9 that $d_j < a_j$. 
This choice also reflects a trade-off between accurately capturing the states with which we have considerable prior information, i.e. \IA, whilst improving the computational efficiency of the other states over a standard \HSMM formulation.

We assume that the night rest period of a healthy individual is generally between 7 and 8 hours. The parameter of the dwell duration of the \IA state, $\lambda_{\,\IA}$,  is hence assigned a Gamma prior with hyperparameters that reflect mean 90  (i.e. $7.5\times 12$) and variance 36 (i.e. $[0.5\times 12]^{2}$), the latter was chosen to account for some variability amongst people. %that possibly wake up overnight. 
Since we do not have significant prior knowledge on how long people spend in the \MA and \HA states, we assigned  $\lambda_{\,\MA}$ and $\lambda_{\,\HA}$ Gamma priors with mean 24 (i.e. 2 hours) and variance 324 (i.e. $[1.5\times 12]^2$) to reflect a higher degree of uncertainty. 
Transition probabilities from state $\IA$, $\bm{\pi}_{\IA}$, are specified as Dirichlet with equal prior probability of switching to any of the active states \MA or \HA. On the other hand, active states usually alternate between each other more frequently than with \IA \citep{huang2018hidden}, and therefore 
we set the prior for $\bm{\pi}_{\MA}$ so that transitions from \MA to \HA are four times more likely 
than switching from \MA to \IA (a similar argument can be made for $\bm{\pi}_{\HA}$). Finally, the inverse of dispersion parameters $\rho_j^{-1}$ were given Gamma$\,(2, 2)$ priors, and the parameters of the Gaussian emissions were assigned $\mu_j \sim$ Normal$\,(\bar{y}, 4)$ and $\sigma^{\,2}_j\, \sim$ Inverse-Gamma$\,(2, 0.5)$, where $\bar{y}$ denotes the sample mean.

% We note that these priors are depending on the data, but only weakly. 

\begin{figure}[htbp] 
	\centering
	%\centerline{\includegraphics[scale = 0.31]{Plots/final_fit_caseStudy.png}}
	\centerline{\includegraphics[width =0.8\linewidth]{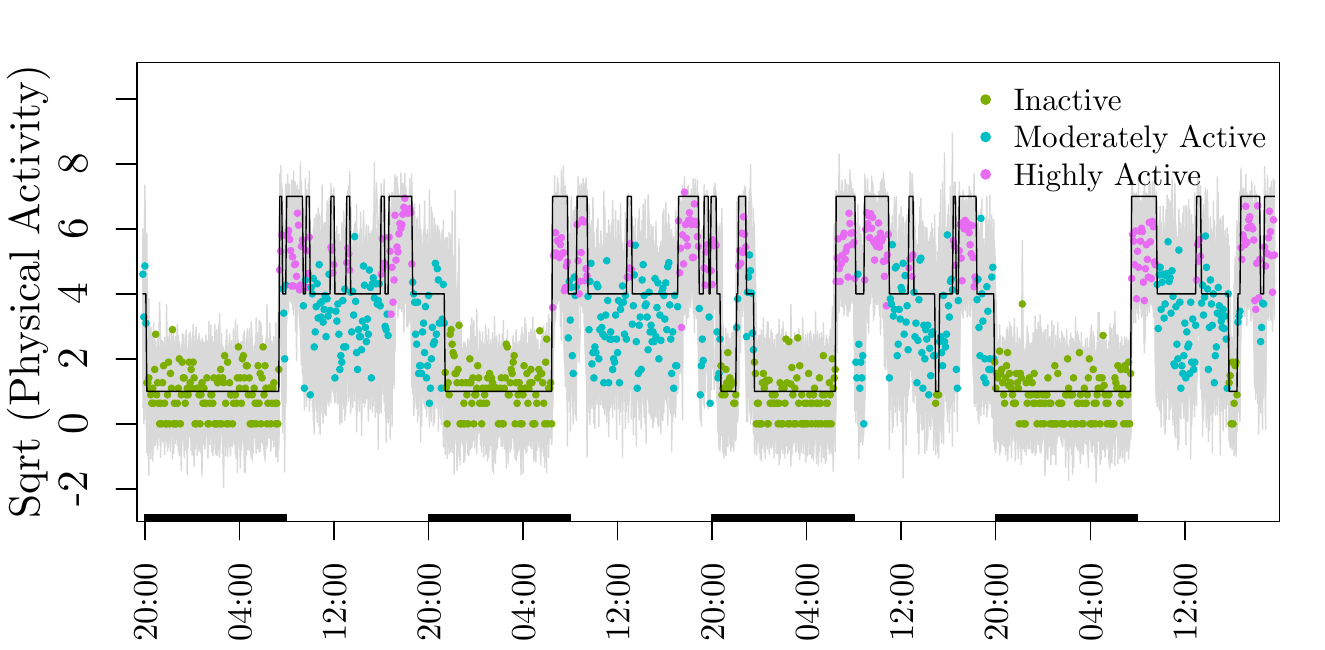}}
	%trim={<left> <lower> <right> <upper>}
	\caption{Square root of the \PA time series along with simulated observations from the fitted model with negative binomial dwell-time. The piecewise horizontal line represents the estimated state sequence. Rectangles on the time axis correspond to periods from 20.00 to 8.00. \IA state happens during night, whereas days are characterized by many switches between \MA and \HA states. This picture is best viewed in color.}
	\label{fig:fit_caseStudy}
\end{figure}

For each proposed model our Bayesian procedure is run for 6,000
iterations, 1,000 of which are discarded as burn-in. Firstly, we consider selecting which of the competing dwell distributions, i.e. the geometric dwell characterising the \HMM and the Poisson and negative binomial \HSMM extensions, is most supported by the observed data. As explained in Section \ref{sec:comparable_dwell}, we specified hyperparameters for these competing models so that the corresponding priors match the means and variances of the informative prior specification given above.  In order to measure the gain of including available prior knowledge into the model, we also investigated 
%predictive performances when specifying 
a weakly informative prior setting (as in Section \ref{sec:ill_example}). Table \ref{table:casestudy_comparison} displays the bridge sampling estimates of the marginal likelihood for the different models and posterior means of the corresponding dwell parameters. It is clear that integrating into the model available prior information improves performance greatly. In addition, modelling dwell durations as either negative binomial or geometric 
%yields superior predictive results 
provides a better approximation to the data 
compared to a Poisson model. 
%Furthermore, the Bayes factor  $7.92$ (i.e. $\exp\{-1633.25 + 1635.32\})$ suggests that there is \textit{substantial evidence} \citep{kass1995bayes} in favour of the $\HSMM$ with negative binomial durations in comparison to a standard \HMM. 
Furthermore, the Bayes factor  $18.36$ (i.e. $\exp\{-1632.42 + 1635.33\})$ suggests that there is \textit{strong evidence} \citep{kass1995bayes} in favour of the $\HSMM$ with negative binomial durations in comparison to a standard \HMM.
This is also reflected by the estimated posterior means of the parameters $\rho_j$ which differ from one, hence showing some departure from geometric dwell durations. These `dispersion' parameters are smaller than one for the \IA and \MA states indicating a larger fitted  variance of the dwell times under the negative binomial \HSMM than the geometric \HMM. \color{black} Combined with their estimated means, this may explain the improved performance of the negative binomial dwell model over the \HMM.  The increased variance allows the time series to better capture the short transitions to \IA states seen in the fitted model (Figure \ref{fig:fit_caseStudy}). This also explains why the Poisson \HSMM performs poorly for this dataset; the fitted Poisson dwell distribution for the \IA state can be seen to have a much smaller variance than the geometric and negative binomial alternatives. Plots comparing the posterior predictive dwell time for the \IA, \MA, and \HA states estimated under the three proposed dwell distributions are provided in the Supplementary Material. Future work could consider more complex dwell distributions to reflect the different patterns of human sleep. For example, a natural extension to the results presented here could be to look at whether a two-component mixture distribution (e.g. Poisson) can aid in better capturing the short excursions to the \IA seen in Figure \ref{fig:fit_caseStudy}.  In the Supplementary Material, we have further investigated the different state classifications provided by the optimal
proposed model (using negative binomial durations) with respect to Poisson and geometric dwells.

\begin{figure}[htbp] 
	\centering
	\includegraphics[width =0.39\linewidth]{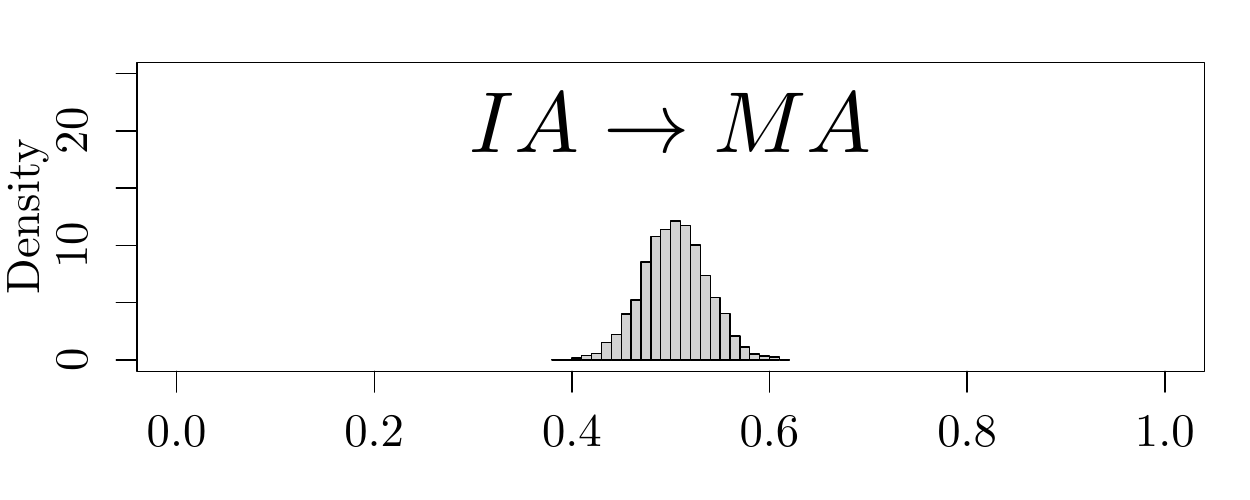}
	\includegraphics[width =0.39\linewidth]{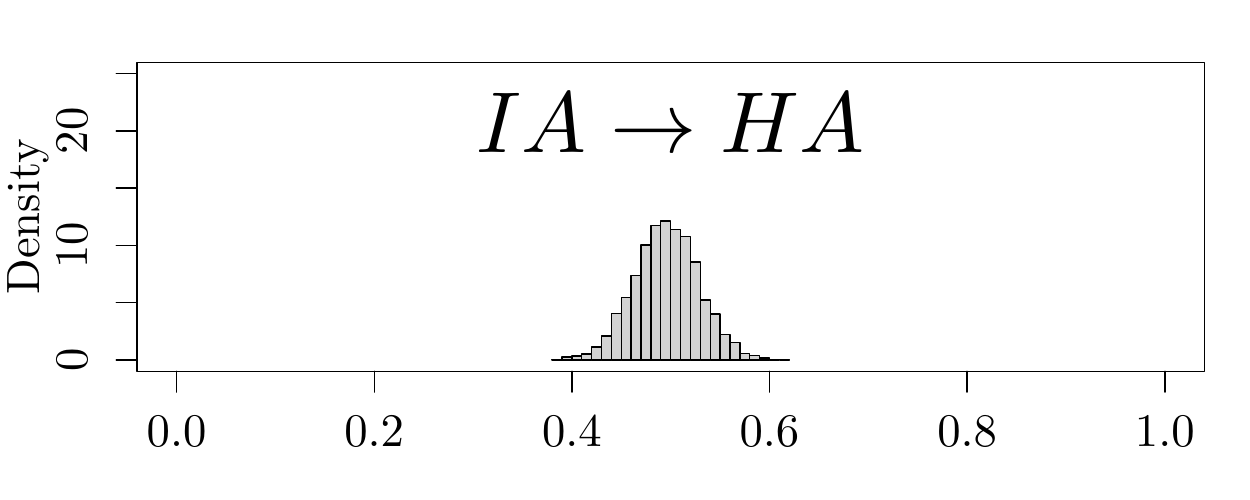}
	\includegraphics[width =0.39\linewidth]{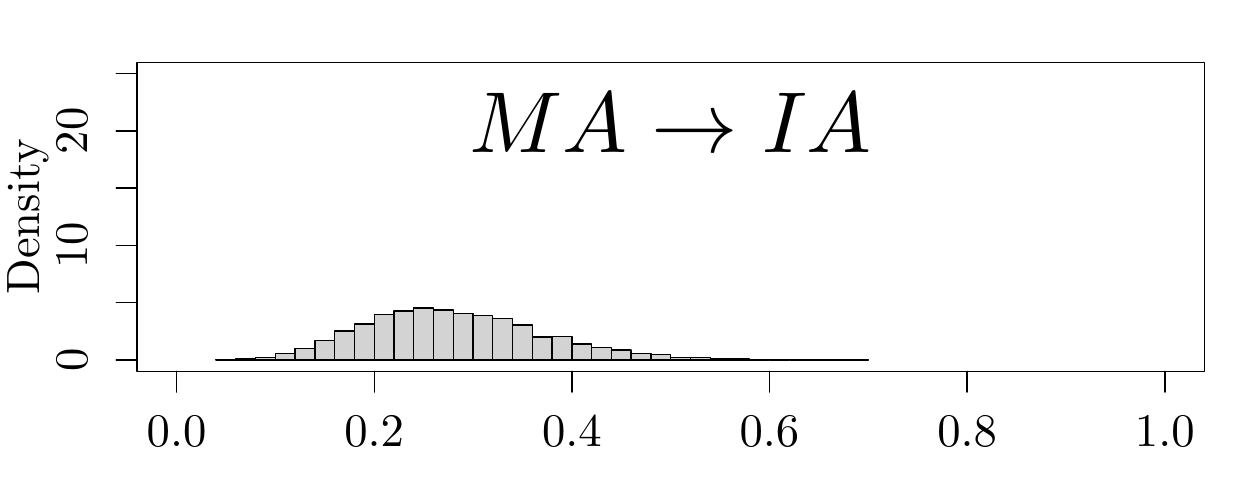}
	\includegraphics[width =0.39\linewidth]{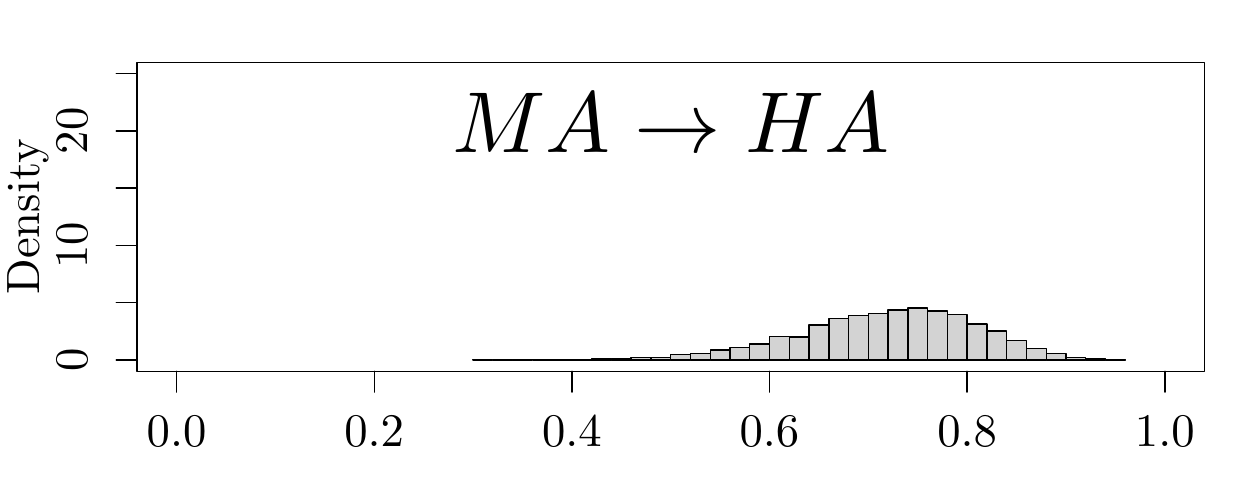}
	\includegraphics[width =0.39\linewidth]{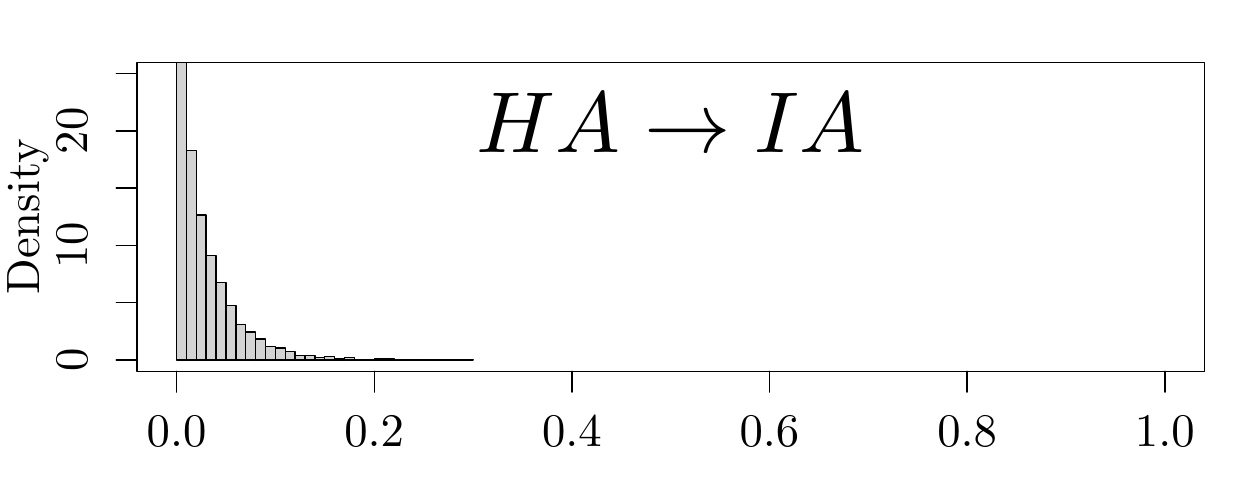}
	\includegraphics[width =0.39\linewidth]{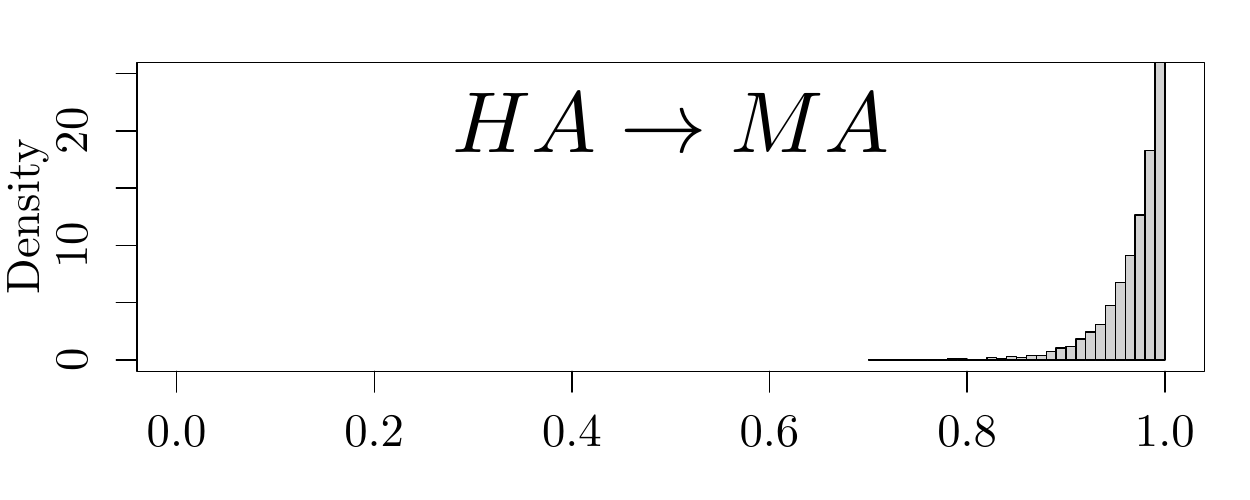}
	\caption{Estimated posterior density histograms of the transition probabilities between \IA, \MA, and \HA states from the proposed \HSMM with negative binomial dwell-times.}
	\label{fig:transitions_caseStudy}
\end{figure}

Posterior means of the emission parameters were $y_t |_{\IA} \sim$ Normal(0.93, 0.47),  $y_t |_{\MA} \sim$ Normal(3.17, 1.28) and $y_t |_{\HA} \sim$ Normal(5.38, 0.54). 
%We observe that the $\IA$ state is characterized by low values and the \MA state has a larger variance than the other two telemetric patterns. Posterior means of dwell parameters are provided in Table \ref{table:casestudy_comparison}, showing an average sleeping behavior of about 7 hours and a half, for this individual. 
The $\IA$ state naturally corresponds to the state with the lowest mean activity and the \MA state appears to have largest variance in activity levels. Posterior means of the dwell parameters in Table \ref{table:casestudy_comparison} show that this individual sleeps an average of 7 and a half hours per night. 
\color{black}
In Figure \ref{fig:transitions_caseStudy}, we display posterior histograms of the transition probabilities between different states. There appears to be high chances of switching between active states, since the posterior means for $\pi_{\HA \rightarrow \MA}$ and $\pi_{\MA \rightarrow \HA}$ are close to one, though the latter exhibits larger variance.  Additionally, the posterior probability of transitioning from \HA to \IA is very close to zero,  which is reasonable considering that it is very unlikely that an individual would go to sleep straight after having performed intense physical activity. Figure \ref{fig:fit_caseStudy} shows the transformed time series as well as simulated data from the predictive distribution, and the estimated hidden state sequence using the Viterbi algorithm. It can be seen that the \IA state occurs during the night whereas days are characterized by many switches between the \MA and \HA states. Our results are in agreement with \citet{huang2018hidden}.

\begin{table}[htbp]
\resizebox{\columnwidth}{!}{%
\begin{tabular}{lccccccc}
			\hline \\[-0.9em]
                  & log-marg lik & $\lambda_{\,\IA}$                                                          & $\lambda_{\,\MA}$                                                       & $\lambda_{\,\HA}$                                                      & $\rho_{\,\IA}$                                                     & $\rho_{\,\MA}$                                                       & $\rho_{\,\HA}$                                                       \\ [.1em] \cmidrule{2-8}
Poisson$\ssymbol{2}$    & $-1751.02$ & \begin{tabular}[c]{@{}c@{}}88.32\\ \footnotesize(86.28–89.28)\end{tabular} & \begin{tabular}[c]{@{}c@{}}34.79\\ \footnotesize(29.08–43.02)\end{tabular} & \begin{tabular}[c]{@{}c@{}}18.55\\ \footnotesize(14.45–22.47)\end{tabular} &      -                                                      &  -                                                          &               -                                             \\[1.0em]
Geometric$\ssymbol{2}$ & $-1653.67$ & \begin{tabular}[c]{@{}c@{}}45.57\\ \footnotesize(26.97–74.42)\end{tabular}    & \begin{tabular}[c]{@{}c@{}}10.53\\ \footnotesize(7.49–14.53)\end{tabular}  & \begin{tabular}[c]{@{}c@{}}8.60\\ \footnotesize(6.13–11.94)\end{tabular}  &   -                                                         &  -                                                          &            -                                                \\[1.0em]
Neg-Binom$\ssymbol{2}$ & $-1649.00$ & \begin{tabular}[c]{@{}c@{}}46.25\\ \footnotesize(21.12–88.14\end{tabular}     & \begin{tabular}[c]{@{}c@{}}10.46\\ \footnotesize(6.22–16.94)\end{tabular}  & \begin{tabular}[c]{@{}c@{}}8.37\\ \footnotesize(5.44–12.31)\end{tabular}  & \begin{tabular}[c]{@{}c@{}}0.61\\ \footnotesize(0.29–1.08)\end{tabular} & \begin{tabular}[c]{@{}c@{}}0.61\\ \footnotesize(0.33–0.98)\end{tabular} & \begin{tabular}[c]{@{}c@{}}1.22\\ \footnotesize(0.60–2.26)\end{tabular} \\[1.0em]
Poisson           & $-1732.16$ & \begin{tabular}[c]{@{}c@{}}88.39\\ \footnotesize(86.65–89.27)\end{tabular}  & \begin{tabular}[c]{@{}c@{}}33.61\\ \footnotesize(28.76–40.55)\end{tabular} & \begin{tabular}[c]{@{}c@{}}17.98\\ \footnotesize(14.35–22.04)\end{tabular} &    -                                                        &  -                                                          &        -                                                    \\[1.0em]
Geometric         & $-1635.33$ & \begin{tabular}[c]{@{}c@{}}88.72\\ \footnotesize(79.63–98.70)\end{tabular}    & \begin{tabular}[c]{@{}c@{}}13.42\\ \footnotesize(9.49–18.68)\end{tabular}   & \begin{tabular}[c]{@{}c@{}}10.97\\ \footnotesize(7.91–15.05)\end{tabular} &  -                                                          &  -                                                          &       -                                                     \\[1.0em]
Neg-Binom         & $\bm{-1632.42}$ & \begin{tabular}[c]{@{}c@{}}87.97\\ \footnotesize(78.40–97.75)\end{tabular}     & \begin{tabular}[c]{@{}c@{}}12.07\\ \footnotesize(7.33–18.84)\end{tabular}  & \begin{tabular}[c]{@{}c@{}}9.12\\ \footnotesize(5.99–13.19)\end{tabular}  & \begin{tabular}[c]{@{}c@{}}0.67\\ \footnotesize(0.33–1.19)\end{tabular} & \begin{tabular}[c]{@{}c@{}}0.71\\ \footnotesize(0.36–1.15)\end{tabular} & \begin{tabular}[c]{@{}c@{}}1.25\\  \footnotesize(0.60–2.22)\end{tabular} \\ [1em] \hline
\end{tabular}
}%
\caption{Telemetric activity data. Log-marginal likelihood for different dwell distributions (Poisson, geometric and negative binomial), where  the superscript $\ssymbol{2}$ denotes a weakly informative prior specification. Geometric durations are characterized by their mean dwell length $\lambda_j = 1/(1 - \gamma_{jj})$ where $\gamma_{jj}$ represents the probability of self-transition. Estimated posterior means of the  dwell parameters are reported  with a 90\% credible intervals estimated from the posterior sample.}
\label{table:casestudy_comparison}
\color{black}
\end{table}

%\begin{figure}[htbp] 
%	\centering
%	%\centerline{\includegraphics[scale = 0.31]{Plots/final_fit_caseStudy.png}}
%	\centerline{\includegraphics[trim = {0.5cm 1.0cm 0.5cm 1.0cm}, clip, scale = 0.31]{Plots/final_fit_caseStudy.png}}
	%trim={<left> <lower> <right> <upper>}
%	\caption{Square root of the \PA time series along with simulated observations from our fitted model with negative binomial dwell-time, where the piecewise horizontal line represents the estimated state sequence. Rectangles on the time axis correspond to periods from 20.00 to 8.00. \IA state happens during night, whereas days are characterized by many switches between \MA and \HA states. This picture is best viewed in color.}
%	\label{fig:fit_caseStudy}
%\end{figure}

%\begin{figure}[htbp] 
%	\centering
%	\centerline{\includegraphics[trim = {0.5cm .5cm 0.5cm 0.5cm}, clip, scale = 0.30]{Plots/transitions_caseStudy.png}}
%	\caption{Estimated posterior density histograms of the transition probabilities between \IA, \MA, and \HA states, where estimation was carried out using our proposed \HSMM with negative binomial dwell-time.}
%	\label{fig:transitions_caseStudy}
%\end{figure}

%\newpage
%\subsection{Harmonic Emissions Density} 
\subsection{Harmonic Emissions} 
\label{sec:harmonic_emissions}

%\markf{Thoughts on moving this to Section 2, but keep the results here? } \jack{I guess the model goes in Section 2 and the Inference in Section 3}

%\jack{The flexibility provided formulating the HSMM as an HMM on an extended state space allows... in addition to standard Gaussian emission in each state}

%Considering that the transition probabilities of the latent Markov process of \PA recordings may be influenced by a circadian oscillator,
%\citet{huang2018hidden} further improved upon a standard Gaussian \HMM by developing an extended model where the structure of the state transition dynamics are augmented by expressing the transition probabilities as a logistic function of the circadian periodicity (24 hours). 

\citet{huang2018hidden} further extended the standard Gaussian \HMM for the \PA recordings by allowing the state transition dynamics to depend on body's circadian periodicity (24 hours).
\color{black}
In a similar vein, we investigate the inclusion of spectral information within the emission density, and study how this affects the \HMM and \HSMM models considered in the previous section.  Specifically, we consider that the observations are generated from state-specific \textit{harmonic emissions}  of the form $y_t \, | \,  z_{\,t} = j \, \sim \, \mathcal{N} \, ( \, \mu_j (t), \sigma^2_j)$, with oscillatory mean  defined as
\begin{equation}
    \mu_j (t) = \beta^{(0)}_j + \beta^{(1)}_j \cos (2\pi\hat{\omega} t) + \beta^{(2)}_j \sin (2\pi\hat{\omega} t).
\label{eq:harmonic_emission}
\end{equation}
This emission density is hence expressed as a sum of a sine and a  cosine (weighted by the linear coefficients $\beta_{j}^{(1)}$ and $\beta_{j}^{(2)}$)  oscillating at frequency $\hat{\omega}$, plus a  state-specific intercept $\beta_{j}^{(0)}$. While \citet{huang2018hidden} choose a priori the 24-hour  periodicity included in the basis function, in our study we estimate this directly from the data. The next section describes our approach for identifying the frequency $\hat{\omega}$ driving the overall variation in the \PA time series.

% Note that  Eq. \eqref{eq:harmonic_emission} assumes that the frequency $\hat{\omega}$ is fixed and does not change across different states  \benni{why does it make sense to have a fixed $\hat{\omega}$?}.

\subsubsection{Identifying the Periodicity} 
%\subsection{Inference for the Periodicity of the Harmonic Emissions} 

%\jack{The additional parameters of the harmonic Emission...}

%Here our main objective is to identify the frequency $\hat{\omega}$ to incorporate as a circadian covariate in the Gaussian emissions, which we define as the posterior mean of the frequency $\omega$ under the periodic model \eqref{eq:periodic_model} defined below, 
We define $\hat{\omega}$ as the posterior mean of the frequency $\omega$ under the periodic model in Eq. \eqref{eq:periodic_model} defined below, 
i.e. $ \hat{\omega} :=  \mathbb{E} \, ( \omega \, | 
\, \bm{y}, \bm{\beta}, \sigma^2)$, with $\bm{\beta} = (\beta^{(1)}, \beta^{(2)})$. 
In this preliminary step to the proposed model with harmonic emissions (Eq. \ref{eq:harmonic_emission}), we  first assume the data to be generated by the following stationary periodic process 
\begin{equation} y_t = \beta^{(1)} \cos (2\pi \omega t) + \beta^{(2)} \sin (2\pi\omega) + \varepsilon_t, \quad \varepsilon_t \sim \mathcal{N}(0, \sigma^2_{\omega}), \qquad t=1, \dots, T, 
\label{eq:periodic_model}
\end{equation} where we have developed a Metropolis-within-Gibbs sampler to obtain samples from the posterior distribution of the frequency 
\begin{equation} \label{posterior_omega}
p \, (\omega \, | \, \bm{\beta},  \, \sigma^2, \, \bm{y} ) \propto \exp \Bigg[ -\dfrac{1}{2\sigma^2} \sum_{ t } \Big\{ y_t - \beta^{(1)} \cos (2\pi \omega t) - \beta^{(2)} \sin (2\pi\omega) \, \Big\}^{2}  \Bigg] \mathbbm{1}_{\big[ \, \omega \,  \in  \, (0, \, \phi_{\omega}) \big]},
\end{equation}
where $\phi_\omega$ is a pre-specified upper bound for the frequency and may be chosen to reflect  prior information about the value of $\omega$, for example focusing only on low frequencies (e.g. $ 0 < \phi_{\omega} < 0.1 $). Full details of the sampling scheme and our prior choice are provided in the Supplementary Material. This algorithm is similar to the within-model move of the ``segment model'' presented in \citet{hadj2019bayesian, hadj2020spectral}, %but with the difference that in our scenario the number of frequencies is fixed at one.
but with the number of frequencies fixed at one.

%\begin{figure}[htbp]%
%    \centering
%    \subfloat[\centering ]{{\includegraphics[height = 5cm, width=6cm]{Plots/traceplots_omega.pdf} }}%
%     \hskip -2ex
%    \subfloat[\centering]{{\includegraphics[height = 5cm, width=7cm]{Plots/posterior_predictive_omega.pdf} }}%
%    \caption{\textcolor{black}{Identifying the periodicity via the periodic model (5.2). Panel (a) shows the trace plot (after burn-in) of the posterior distribution of the frequency. Panel (b) displays draws from the posterior predictive as well as the estimated periodic signal. In both plots, the red represents posterior mean. }}%
%    \label{fig:example_oscillatory}%
%\end{figure}

\begin{figure}[htbp]%
    \centering
    \subfloat[\centering ]{{\includegraphics[width =0.49\linewidth]{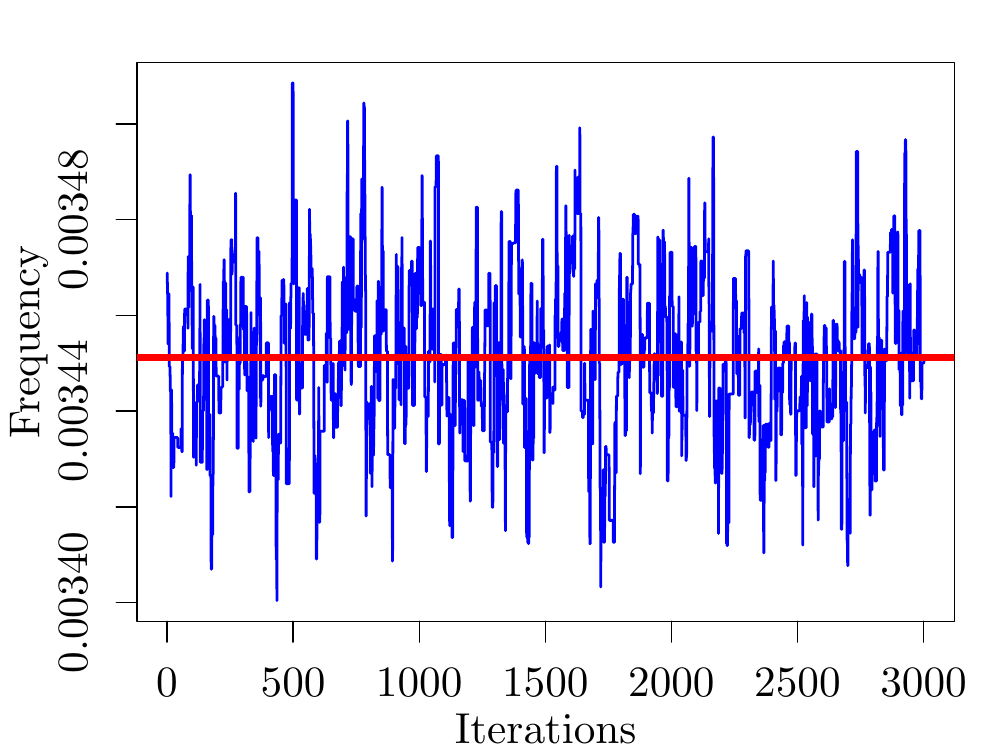} }}%
     \hskip -2ex
    \subfloat[\centering]{{\includegraphics[width =0.49\linewidth]{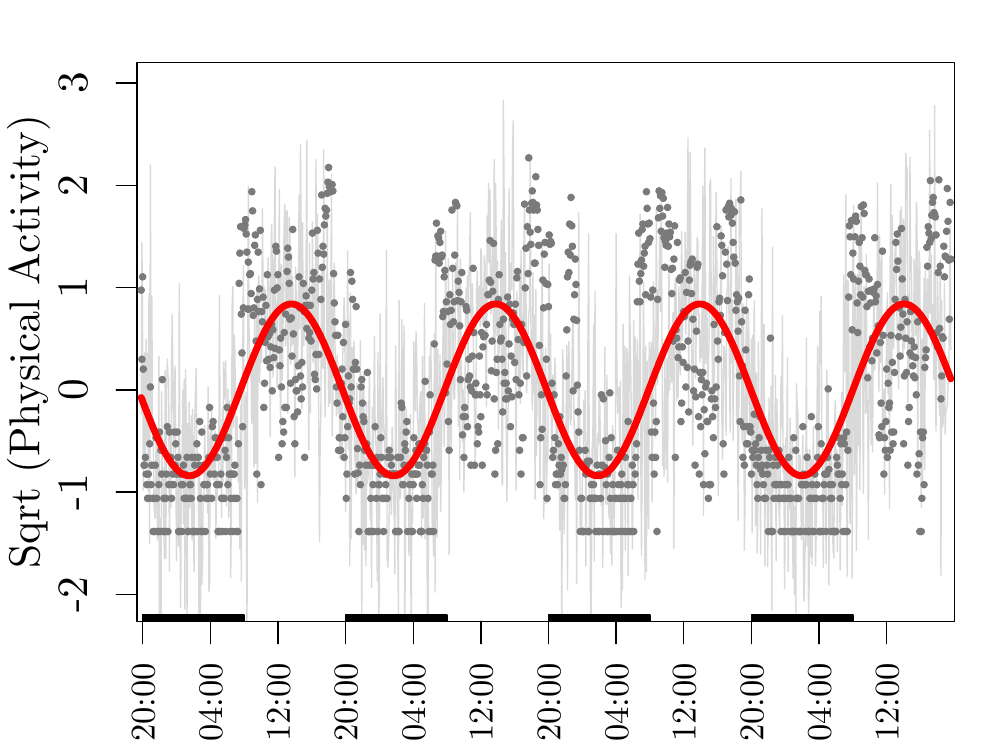} }}%
    \caption{\textcolor{black}{Identifying the periodicity via the periodic model in Eq. \eqref{eq:periodic_model}. Panel (a) shows the trace plot (after burn-in) of the posterior distribution of the frequency. Panel (b) displays draws from the posterior predictive as well as the estimated periodic signal. In both plots, the red represents posterior mean. }}%
    \label{fig:example_oscillatory}%
\end{figure}

We ran the sampler for 5000 iterations using software written in Julia 1.6 which took around 3 seconds on an Intel\textsuperscript{\textregistered} Core\textsuperscript{TM} i5  2 GHz Processor with 16 GB RAM. Figure \ref{fig:example_oscillatory} (a) shows the trace plot (after burn-in) of the posterior sample of the frequency where the acceptance rate (28\%) was roughly tuned to be optimal \citep{roberts2001optimal}. We also highlight in red the posterior mean $\hat{\omega} =  0.003453$. In Figure \ref{fig:example_oscillatory} (b) we display 20 draws from the posterior predictive distribution of the stationary periodic model and the posterior mean of the oscillatory signal. This shows that the model predictions appear to capture some of the structure of the \PA time series. However, there also appears to be temporal structure not captured by the global circadian harmonic. As a result, in the next section we will use the global $\hat{\omega} =  0.003453$ as the circadian covariate for the emissions of the harmonic \HMM and \HSMM (Eq. \ref{eq:harmonic_emission}), allowing the harmonic parameters $(\beta_j^{(0)}, \beta_j^{(1)}, \beta_j^{(2)})$ to vary by state in order to better capture the temporal structure. 

\vspace{0.3cm}

%\subsection{The Harmonic Emission} 
\subsubsection{Results}

Given the point estimate for $\hat{\omega} =  0.003453$, we then applied the \HMM and \HSMM approximations  with Poisson and negative binomial dwells to the \PA time series  (using $K = 3$ states). Our prior specification follows the discussion in Section \ref{sec:application} for $\sigma^2_j$, $\lambda_j$, $\gamma_j$ and $\rho_j$, where appropriate, while the intercept of the harmonic mean model $\beta_j^{(0)}$ is given the same prior as $\mu_j$ from the standard Gaussian emission model. The additional parameters of the harmonic model $\beta_j^{(1)}$ and $\beta_j^{(2)}$ are both assumed a priori $\mathcal{N}(0, 2^2)$.

Table \ref{table:casestudy_comparison_harmonic} (top) provides the log-marginal likelihoods of the different models and posterior mean estimates of the parameter of their dwell-distributions, along with the 90\% credibility intervals provided by their posteriors. It is clear that the marginal likelihood favours the negative binomial dwell distribution, with the standard \HMM (geometric dwell) being the next most favorable. Further, when comparing Table \ref{table:casestudy_comparison_harmonic} with Table \ref{table:casestudy_comparison}, we see that the inclusion of harmonic emissions results in an increase of the marginal likelihood by a factor ranging between 6 and 7 on the log-scale for all dwell distributions, thus supporting its integration in our model.

% Furthermore, when comparing Table \ref{table:casestudy_comparison_harmonic} with Table \ref{table:casestudy_comparison} we see that for all the dwell distributions including the harmonic emission increased to marginal likelihood by a factor of between 6 and 7 on the log-scale, thus supporting its inclusion in the model.

\begin{table}[htbp]
\resizebox{\columnwidth}{!}{%
\begin{tabular}{lccccccc}
			\hline \\[-0.9em]
                  & log-marg lik & $\lambda_{\,\IA}$                                                          & $\lambda_{\,\MA}$                                                       & $\lambda_{\,\HA}$                                                      & $\rho_{\,\IA}$                                                     & $\rho_{\,\MA}$                                                       & $\rho_{\,\HA}$                                                       \\ [.1em] \cmidrule{2-8}
Poisson           & $-1727.24$ & \begin{tabular}[c]{@{}c@{}}88.29\\ \footnotesize(86.42–89.25)\end{tabular}  & \begin{tabular}[c]{@{}c@{}}44.68\\ \footnotesize(41.80–47.57)\end{tabular} & \begin{tabular}[c]{@{}c@{}}21.62\\ \footnotesize(18.52–25.05)\end{tabular} &    -                                                        &  -                                                          &        -                                                    \\[1.0em]
Geometric         & $-1629.40$ & \begin{tabular}[c]{@{}c@{}}88.24\\ \footnotesize(79.08–98.05)\end{tabular}    & \begin{tabular}[c]{@{}c@{}}15.53\\ \footnotesize(10.19–23.08)\end{tabular}   & \begin{tabular}[c]{@{}c@{}}12.15\\ \footnotesize(8.36–17.61)\end{tabular} &  -                                                          &  -                                                          &       -                                                     \\[1.0em]
Neg-Binom         & $\bm{-1625.61}$ & \begin{tabular}[c]{@{}c@{}}87.54\\ \footnotesize(77.66–97.34)\end{tabular}     & \begin{tabular}[c]{@{}c@{}}14.32\\ \footnotesize(7.82–24.02)\end{tabular}  & \begin{tabular}[c]{@{}c@{}}10.67\\ \footnotesize(6.44–16.20)\end{tabular}  & \begin{tabular}[c]{@{}c@{}}0.65\\ \footnotesize(0.31–1.17)\end{tabular} & \begin{tabular}[c]{@{}c@{}}0.64\\ \footnotesize(0.33–1.13)\end{tabular} & \begin{tabular}[c]{@{}c@{}}1.7\\  \footnotesize(0.63–3.88)\end{tabular} \\ [1em] \hline
\end{tabular}
}\\ [.1em]%
\resizebox{\columnwidth}{!}{%
\begin{tabular}{lccccccccc}
\hline \\[-0.9em]
  & \multicolumn{3}{c}{\IA}\ & \multicolumn{3}{c}{\MA} & \multicolumn{3}{c}{\HA}
  %\\ [.1em] \cmidrule{2-8}
  \\ [.1em] \hline \\[-0.9em]
  & $\beta^{(0)}$ & $\beta^{(1)}$ & $\beta^{(2)}$ & $\beta^{(0)}$ & $\beta^{(1)}$ & $\beta^{(2)}$ & $\beta^{(0)}$ & $\beta^{(1)}$ & $\beta^{(2)}$
  \\ [.1em] \hline \\[-0.9em]
  Gaussian & \begin{tabular}[c]{@{}c@{}}0.93\\ \footnotesize(0.88-0.98)\end{tabular}  & - & - &  \begin{tabular}[c]{@{}c@{}}3.18\\ \footnotesize(3.03-3.33)\end{tabular} & - & - & \begin{tabular}[c]{@{}c@{}}5.38\\ \footnotesize(5.27-5.51)\end{tabular} & - & -\\[1.0em] \\
  Harmonic & \begin{tabular}[c]{@{}c@{}}1.36\\ \footnotesize(1.26-1.46)\end{tabular}  & \begin{tabular}[c]{@{}c@{}}0.04\\ \footnotesize(-0.05-0.13)\end{tabular} & \begin{tabular}[c]{@{}c@{}}-0.60\\ \footnotesize(-0.72- -0.47)\end{tabular}  & \begin{tabular}[c]{@{}c@{}}3.32\\ \footnotesize(2.94-3.65)\end{tabular} & \begin{tabular}[c]{@{}c@{}}-0.11\\ \footnotesize(-0.34-0.13)\end{tabular} & \begin{tabular}[c]{@{}c@{}}-0.24\\ \footnotesize(-0.69-0.17)\end{tabular} & \begin{tabular}[c]{@{}c@{}}5.46\\ \footnotesize(5.32-5.60)\end{tabular} & \begin{tabular}[c]{@{}c@{}}0.20\\ \footnotesize(0.07-0.33)\end{tabular} & \begin{tabular}[c]{@{}c@{}}-0.23\\ \footnotesize(-0.61-0.16)\end{tabular}\\ [1em] \hline
\end{tabular}
}%
\caption{\textcolor{black}{Telemetric activity data with harmonic emissions. (Top) Log-marginal likelihood for different dwell durations (i.e. Poisson, geometric and negative binomial). Geometric durations are characterized by their mean dwell length $\lambda_j = 1/(1 - \gamma_{jj})$ where $\gamma_{jj}$ represents the probability of self-transition. (Bottom) Parameters of the mean of the Gaussian and harmonic emission distributions under the selected negative binomial dwell distribution. Estimated posterior means of the parameters are reported with a 90\% credible intervals estimated from the posterior sample.}} 
\label{table:casestudy_comparison_harmonic}
\color{black}
\end{table}

Following the selection of the negative binomial dwell distribution for both the standard Gaussian and harmonic emission models, Table \ref{table:casestudy_comparison_harmonic} (bottom) provides the posterior mean values for the parameters of these emission distributions, along with the 90\% credibility intervals provided by the posterior. These results show that even with a global estimate for the periodicity, there are differences between the estimated parameters in each state, supporting the combination of the periodic time-series model with a hidden state model. Furthermore, there are clear differences between the estimated emissions of the harmonic model compared with the estimated Gaussian emissions in the standard model (where $\beta^{(0)}_j = \mu_j$ and $\beta^{(1)}_j$ and $\beta^{(2)}_j$ were both 0). In particular, the intercept $\beta^{(0)}_{\,\IA}$ in the \IA state differs non-negligibly when using the harmonic model instead of the standard Gaussian, as do $\beta^{(2)}_{\,\IA}$ and $\beta^{(1)}_{\,\HA}$, whose 90\% credibility intervals do not cover 0. This all supports the selection of the harmonic model over the standard Gaussian emissions.

%\newpage
\section{Concluding Summaries}

% \jack{OLD:}
% We presented a Bayesian model for analyzing time series data based on the \HSMM approximation introduced by \citet{langrock2011hidden} in which a special structure of the transition matrix is embedded to model the state duration distributions. We showed the advantages of choosing a  Bayesian paradigm over its frequentist counterpart in terms of incorporation of prior information, quantification of uncertainty, model selection, and forecasting. The proposed approach allows for the development of highly flexible and interpretable models that incorporate available prior information on state durations. Our methodology is applied to physical activity data collected from a wearable sensing device, where we successfully described the probabilistic dynamics governing the transitions between different activity patterns during the day as well as characterizing the sleep duration overnight. Future work will address extending our methodology to multivariate time series as well as including covariates. \benni{add a better discussion}.

We presented a Bayesian model for analyzing time series data based on an \HSMM formulation with the goal of analyzing physical activity data collected from wearable sensing devices. We facilitate the computational feasibility of Bayesian inference for \HSMMs via the likelihood approximation introduced by \citet{langrock2011hidden}, in which a special structure of the transition matrix is embedded to model the state duration distributions. We utilize the \stan{} modeling language and deploy a sparse matrix formulation to further leverage the efficiency of the approximate likelihood. We showed the advantages of choosing a  Bayesian paradigm over its frequentist counterpart in terms of incorporation of prior information, quantification of uncertainty, model selection, and forecasting. We additionally demonstrated the ability of the \HSMM approximation to drastically reduce the computational burden of the Bayesian inference (for example reducing the time for inference on $T = 5000$ observations from $>3$ days to $<2$ hours), whilst incurring negligible statistical error. The proposed approach allows for the efficient implementation of highly flexible and interpretable models that incorporate available prior information on state durations.  An avenue not explored in the current paper is how our model compares to particle filtering methods. For example, a referee suggested that an algorithm sampling the filtering distribution using an adaptation of the sequential Monte Carlo (SMC) sampler of \cite{yildirim2013online} inside one of the two particle \MCMC algorithms of \cite{whiteley2009particle} could prove competitive for \HSMM inference. Further work could define, implement and compare such an approach to ours. 

% We presented a Bayesian model for analyzing time series data based on a \HSMM formulation with the goal of analysing physical activity data collected from wearable sensing devices. Both Bayesian and \HSMM inference are generally plagued by their computational complexity. We facilitate the computational feasibility of Bayesian inference for \HSMMs by combining the \HSMM approximation introduced by \citet{langrock2011hidden}, in which a special structure of the transition matrix is embedded to model the state duration distributions, with the \stan{} modeling language and deploy a sparse matrix formulation to further leverage the efficiency of the approximate likelihood. We showed the advantages of choosing a  Bayesian paradigm over its frequentist counterpart in terms of incorporation of prior information, quantification of uncertainty, model selection, and forecasting, whilst demonstrating the ability of the \HSMM approximation to drastically reduce the computational burden of the Bayesian inference (for example reducing the time for inference on $T = 5000$ observations from $>3$ days to $<2$ hours), whilst incurring negligible statistical error. The proposed approach allows for the efficient implementation of highly flexible and interpretable models that incorporate available prior information on state durations.

The analysis of physical activity data demonstrated that our model was able to learn the probabilistic dynamics governing the transitions between different activity patterns during the day as well as characterizing the sleep duration overnight. We were also able to illustrate the flexibility of the proposed model by adding harmonic covariates to the emission distribution, extending further the analysis of \cite{huang2018hidden}. Future work will investigate the further inclusion of covariates into these time series models as well as computationally and statistically efficient approaches for conducting variable selection among these \citep{george1993variable, rossell2017nonlocal}. We will also consider extending our methodology to account for higher-dimensional multivariate time series, where computational tractability is further challenging.

%\jack{add discussion. Make two computationally intensive procedures feasible whilst also improving estimation, non-trivial sparse implementation, facilitated cool stuff for the applied example, and cool results}

%\jack{Maybe this sentence is better in the conclusion as to future work, and maybe we point out some reference, spike, and slab or non-local priors blabla, what do you think? Here, we have used the coverage of the 90\% credibility intervals as a crude proxy for the `selection' of a variable. One could make this more rigorous by further doing variable selection for the $\beta_j$'s.  }

%In particular,  we seek to integrate into our applied analysis the skin-surface temperature time series that is collected from the same wearable device that produces physical activity counts \citep{hadj2019bayesian}. We believe that investigating the analysis of associations between skin temperature and physical activity for multiple subjects would possibly improve the understanding of the biological implications underlying the rest-activity rhythms of individuals as they go about their daily lives.

\color{black}

%\section*{Supplementary Material} 
%Supplementary materials are available and include further details about dwell durations, forecasting functions, graphs of normal pseudo-residuals, and further analysis of the \PA time series results. Code that implements the methodology is available as online supplemental material (see also \url{https://github.com/Beniamino92/BayesianApproxHSMM}).%{https://github.com/Beniamino92/BayesianApproxHSMM})

% \begin{supplement}
% % \sname{Supplement A}\label{suppA} 
% % \stitle{\benni{Title of the Supplement A}}

% We provide supplemental material to the manuscript. \slink[url]{https://github.com/Beniamino92/BayesianApproxHSMM}

% \end{supplement}

%\bibliographystyle{ba}
%\bibliography{bib/biblio}
\bibliography{biblio}

\section*{Acknowledgements} 
The authors would like to thank the Editor, the Referee, the Associate Editor, David Rossell, and Marina Vannucci for their insightful and valuable comments.
%\end{document}

% \begin{supplement}
% % \sname{Supplement A}\label{suppA} 
% % \stitle{\benni{Title of the Supplement A}}

% We provide supplemental material to the manuscript. \slink[url]{https://github.com/Beniamino92/BayesianApproxHSMM}

% \end{supplement}
\newpage

\appendix

\section{Supplementary Material} 

%This document includes supplementary material to the article ``Bayesian Approximations to Hidden Semi-Markov Models for Telemetric Monitoring of Physical Activity''. 
Section \ref{sec:mean_variance_dwell} provides the derivations of the mean and variance of the mean dwell time of an \HMM. Section \ref{sec:forecast} and \ref{sec:pseudo} includes the form of the forecast function and graphs of pseudo residuals that we utilized in our experiments.  In Section \ref{sec:state_classification} we provide further results about our \PA data application by comparing different predictive distributions of the state durations as well as investigating state classification.  Section \ref{sec:periodicity} illustrates the details of our  Metropolis-within-Gibbs sampler to obtain posterior samples of the frequency. Code that implements the methodology is available as online supplemental material (see also \url{https://github.com/Beniamino92/BayesianApproxHSMM}).

\subsection{Mean and Variance of  the Mean Dwell Time in an HMM} \label{sec:mean_variance_dwell}

%  The negative binomial distribution has probability mass function given by
% \begin{equation*}
%     \text{Neg-Binomial}\, (r \, | \, \lambda, \, \rho) = {r + \rho - 1 \choose r} \bigg(\dfrac{\lambda}{\lambda + \rho}\bigg)^{r-1} \bigg(\dfrac{\rho}{\lambda + \rho}\bigg)^{\rho},
% \end{equation*}a
% where $r$ $\in \mathbb{N}_{>0}$. This configuration includes a location parameter $\lambda$ $\in \mathbb{R}_{>0}$ and a parameter $\rho$ $\in \mathbb{R}_{>0}$ that regulates overdispersion. The mean and variance of this random variable are 
% \begin{equation}
%     \mathbb{E}\,[ \,d\,] = \lambda + 1 \quad \text{and} \quad \text{Var}\,[\,d\,] = \lambda + \dfrac{\lambda^2}{\rho}.
% \end{equation} Note that $\lambda^2/\rho$ is an additional factor to the variance of a Poisson$\,$($\lambda$). While a Poisson$\,$($\lambda$) describes its mean and variance with the same value, the negative binomial allows for further modelling of the precision through an additional factor $\rho$. \benni{need to rewrite it and say who is a random variable, say $d \sim Neg-Binom()$}

Here we provide the derivations of the mean and variance of the mean dwell time of an \HMM as explained in Section \textcolor{blue}{3.1} of the main paper. Consider a standard \HMM  (Eq. \textcolor{blue}{(1.1)} in the manuscript) with transition probabilities $\{ \bm{\gamma}_j \}_{j=1}^{K}$. Let us assume  that $\bm{\gamma}_{j} = (\gamma_{j1}, \ldots, \gamma_{jK})\sim \textrm{Dirichelt}\left(v_{j1}, \ldots, v_{jK}\right)$ and thus, marginally, $\gamma_{jj}\sim \textrm{Beta}(v_{j}, \beta_j)$, where $v_j := v_{jj}$ and $\beta_j := \sum_{i\neq j} v_{ji}$. The dwell duration in any state follows a geometric distribution with failure probability $1-\gamma_{jj}$, and hence the mean dwell time is given by $\tau_j := \frac{1}{1-\gamma_{jj}}$. As a result, the first and second moments of  the mean dwell time of an \HMM in state $j$   are given by \begin{align}
\mathbb{E}\left[\tau_j\right]& = \int_{0}^{1} \frac{1}{1-\gamma_{jj}}\frac{\Gamma\left(v_j + \beta_j\right)}{\Gamma\left(v_j\right)\Gamma\left(\beta_j\right)}\left(\gamma_{jj}\right)^{v_j-1}\left(1-\gamma_{jj}\right)^{\beta_j-1}d\gamma_{jj}\nonumber\\
&= \int_{0}^{1} \frac{\Gamma\left(v_j + \beta_j\right)}{\Gamma\left(v_j\right)\Gamma\left(\beta_j\right)}\left(\gamma_{jj}\right)^{v_j-1}\left(1-\gamma_{jj}\right)^{\beta_j - 1 - 1}d\gamma_{jj}\nonumber\\
&= \frac{\Gamma\left(v_j + \beta_j\right)\Gamma\left(\beta_j - 1\right)}{\Gamma\left(v_j + \beta_j - 1\right)\Gamma\left(\beta_j\right)}\int_{0}^{1} \frac{\Gamma\left(v_j + \beta_j - 1\right)}{\Gamma\left(v_j\right)\Gamma\left(\beta_j - 1\right)}\left(\gamma_{jj}\right)^{v_j-1}\left(1-\gamma_{jj}\right)^{\beta_j - 1 -1 }d\gamma_{jj}\nonumber\\
&= \frac{v_j + \beta_j -1}{\beta_j - 1}
\end{align}
\begin{align}
\mathbb{E}\left[\tau_j^2\right]& = \int_{0}^{1} \frac{1}{\left(1-\gamma_{jj}\right)^2}\frac{\Gamma\left(v_j + \beta_j\right)}{\Gamma\left(v_j\right)\Gamma\left(\beta_j\right)}\left(\gamma_{jj}\right)^{v_j-1}\left(1-\gamma_{jj}\right)^{\beta_j-1}d\gamma_{jj}\nonumber\\
&= \int_{0}^{1} \frac{\Gamma\left(v_j + \beta_j\right)}{\Gamma\left(v_j\right)\Gamma\left(v_j\right)}\left(\gamma_{jj}\right)^{v_j-1}\left(1-\gamma_{jj}\right)^{\beta_j - 2 -1}d\gamma_{jj}\nonumber\\
&= \frac{\Gamma\left(v_j + \beta_j\right)\Gamma\left(\beta_j - 2\right)}{\Gamma\left(v_j + \beta_j - 2\right)\Gamma\left(\beta_j\right)}\int_{0}^{1} \frac{\Gamma\left(v_j + \beta_j - 2\right)}{\Gamma\left(v_j\right)\Gamma\left(\beta_j - 2\right)}\left(\gamma_{jj}\right)^{v_j-1}\left(1-\gamma_{jj}\right)^{\beta_j - 2 -1}d\gamma_{jj}\nonumber\\
&= \frac{(v_j + \beta_j -1)(v_j + \beta_j - 2)}{(\beta_j - 1)(\beta_j - 2)}\\
\text{Var}\,[\tau_j] &= \frac{(v_j + \beta_j -1)(v_j + \beta_j - 2)}{(\beta_j - 1)(\beta_j - 2)} - \left(\frac{v_j + \beta_j -1}{\beta_j - 1}\right)^2
\end{align}

\subsection{Forecast Density Function} \label{sec:forecast}
Here, we provide the explicit form of the forecasting density $p \,(\tilde{y}_h \, | \, \hat{\bm{\eta}})$ that we used to evaluate predictive performances on a test set $\tilde{\bm{y}} = (\tilde{y}_{\,1}, \, \ldots, \tilde{y}_H)$, with $\tilde{y}_{\,h} = y_{\, T +h}, \hspace{0.1cm} h = 1, \dots, H$, and $H \in \mathbb{N}_{>0}$ denoting the forecast horizon. As in  \citet{zucchini2017hidden}, we express the forecast distribution in the following form
\begin{equation*}
    p \,(\tilde{y}_h \, | \, \hat{\bm{\eta}})  = \bm{\xi}^{'} \,  \bm{\Phi}^{\,h} \, \bm{P}( \tilde{y}_h) \, \bm{1},
\end{equation*} where \begin{equation*}
    \bm{\xi} = \dfrac{\bm{\alpha}(T)}{\mathscr{L}\,  ( \bm{y} \, | \, \bm{\eta})}, 
\end{equation*}
and \begin{equation}
    \mathscr{L}\,  ( \bm{y} \, | \, \bm{\eta}) = \bm{\alpha}(T)^{'} \bm{1}. 
\end{equation} Here, $\bm{P}( y) $ and $\mathscr{L}\,  ( \bm{y} \, | \, \bm{\eta})$ are defined as in Section \textcolor{blue}{3} of the manuscript, and $\bm{1}$ is an $\bar{A}$-dimensional column vector of ones. The vector of  forward-messages $\bm{\alpha}(t) = (\alpha_{1t}, \dots, \alpha_{\bar{A}t})$ can be computed recursively as
\begin{equation*}
    \bm{\alpha}(t+1) = \bm{\alpha}(t) \, \bm{\Phi} \, \bm{P}( y_t), \quad t = 1, \dots T-1.
\end{equation*}

\subsection{Normal Pseudo-Residuals} \label{sec:pseudo}
In order to assess the general goodness of fit of the models that we used in our experiments, we also investigated graphs of  normal pseudo-residuals. Following \citet{zucchini2017hidden}, the normal pseudo-residuals are defined as
\begin{equation}
    r_t = \Psi^{-1} \big[ \,  p \, ( Y_{t} < y_t \, | \, \bm{y}^{(-t)},\, \bm{\eta} ) \, \big],
    \label{eq:psuedo_res}
\end{equation}
where $\Psi$ is the cumulative distribution function of the standard normal distribution and the vector $\bm{y}^{(-t)} = (y_1, \dots, y_{t-1}, y_{t+1}, \dots, y_{T})$ denotes all observations excluding $y_t$. If the model is accurate, $r_t$ is a realization of a standard normal random variable. We provide below index plots of the normal pseudo-residuals, their histograms and quantile-quantile (Q-Q) plots, for our main experiments.

\begin{figure}[htbp] 
	\centering
	\centerline{\includegraphics[height =7cm, width = 17cm]{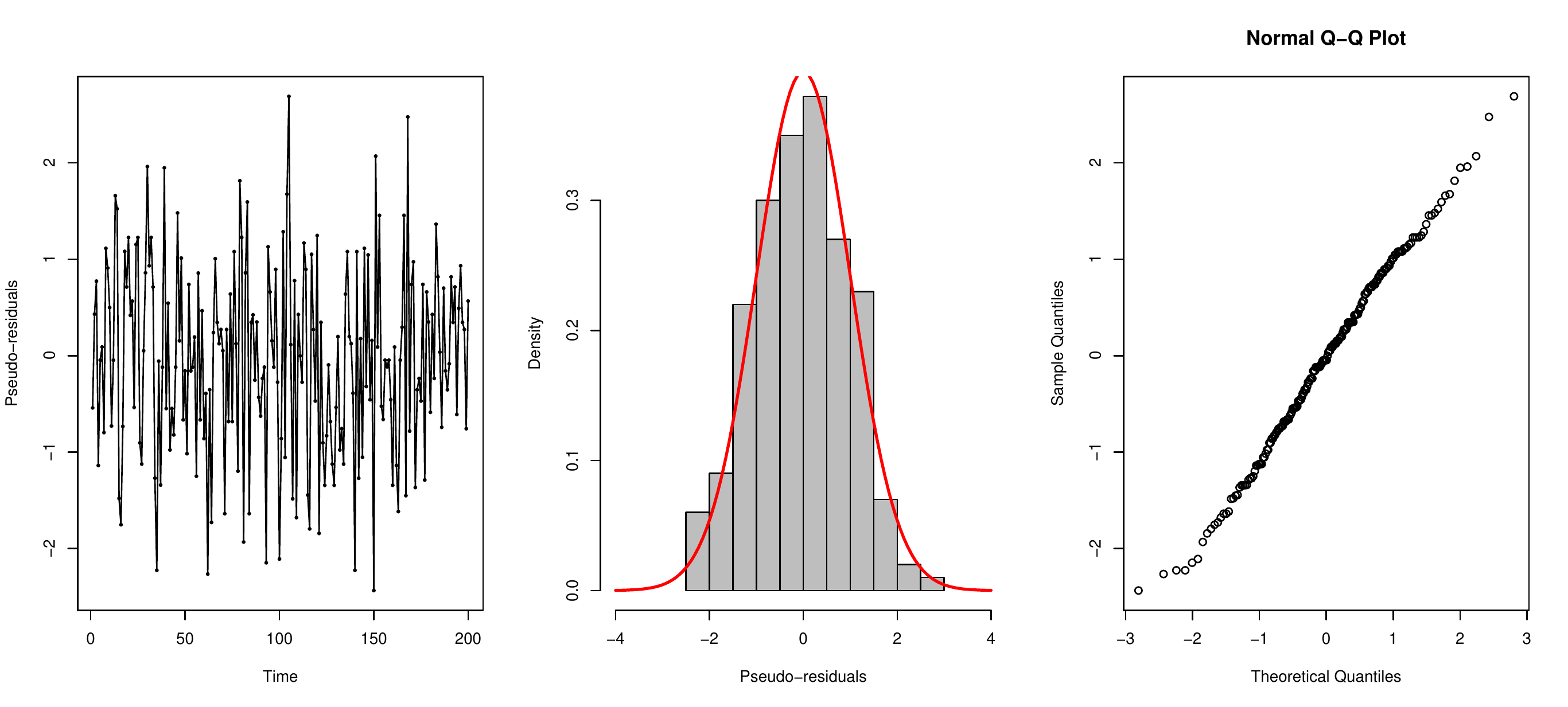}}
	\caption{Illustrative Example. Pseudo residuals: time series, histogram and Q-Q plot.}
	\label{fig:ps_ill_ex}
\end{figure}
\begin{figure}[htbp] 
	\centering
	\centerline{\includegraphics[height =7cm, width = 17cm]{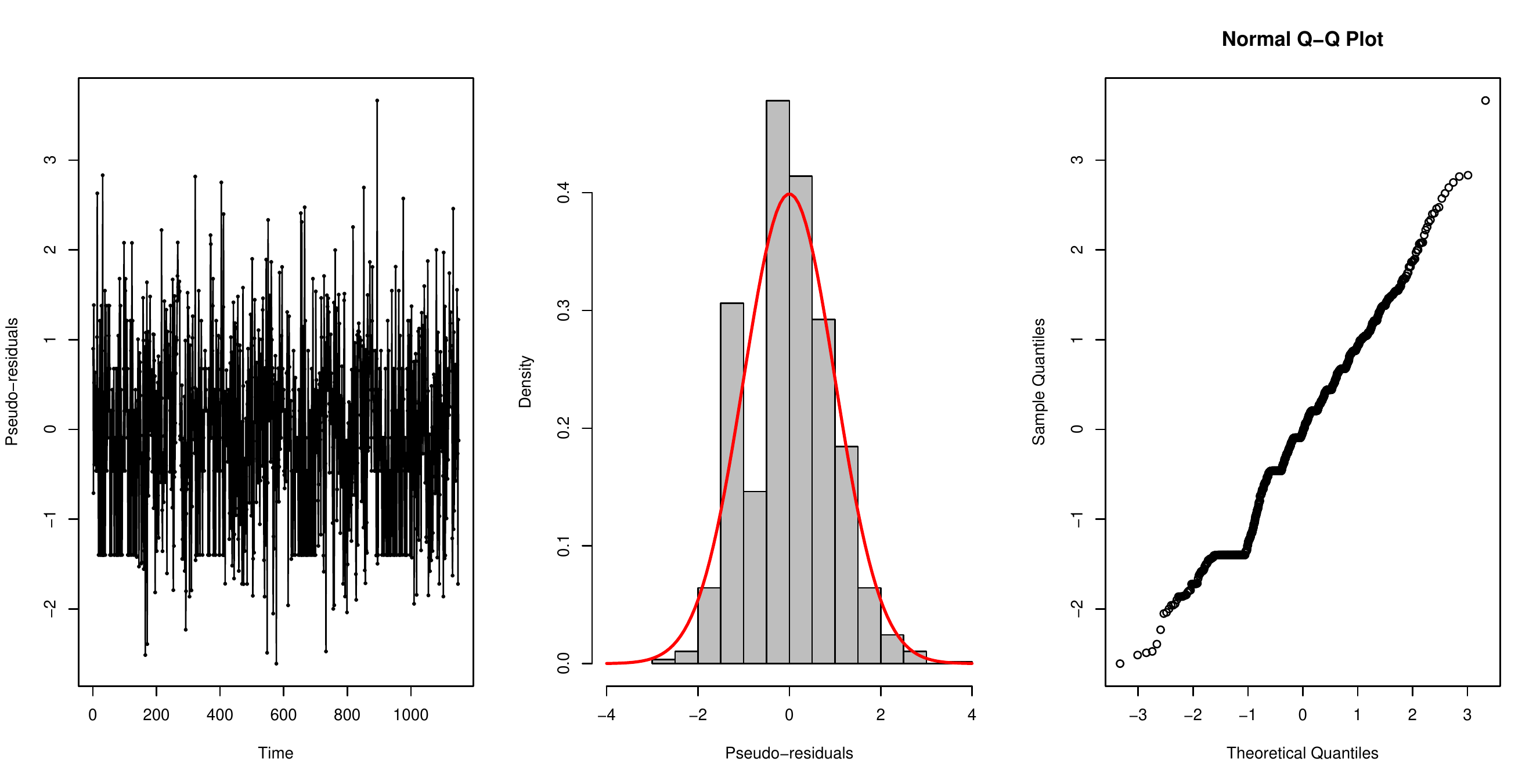}}
	\caption{Telemetric Activity Data. Pseudo residuals: time series, histogram and Q-Q plot.}
	\label{fig:ps_casestudy}
\end{figure}

%\newpage
\subsection{Telemetric Activity Data: Further Results} \label{sec:state_classification}

\subsubsection{Estimated Dwell Distribution}
\begin{figure}[htbp] 
	\centering
	\centerline{\includegraphics[width =0.49\linewidth]{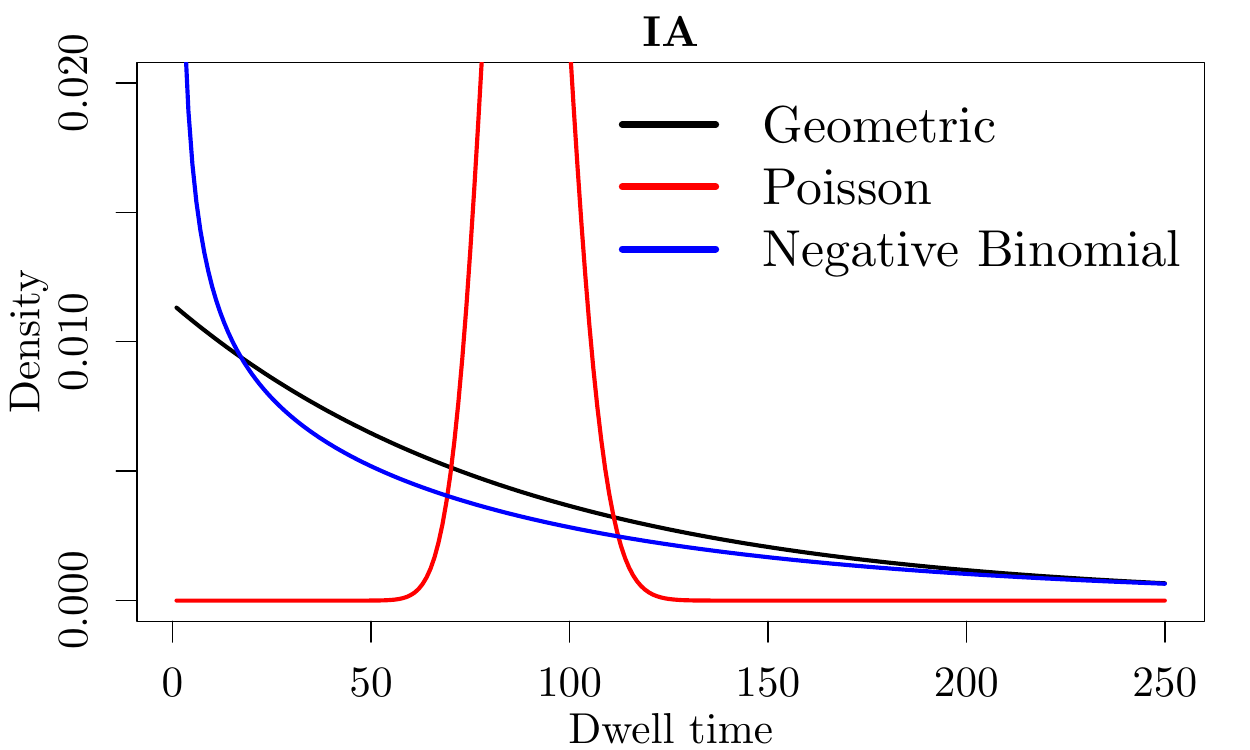}
	            \includegraphics[width =0.49\linewidth]{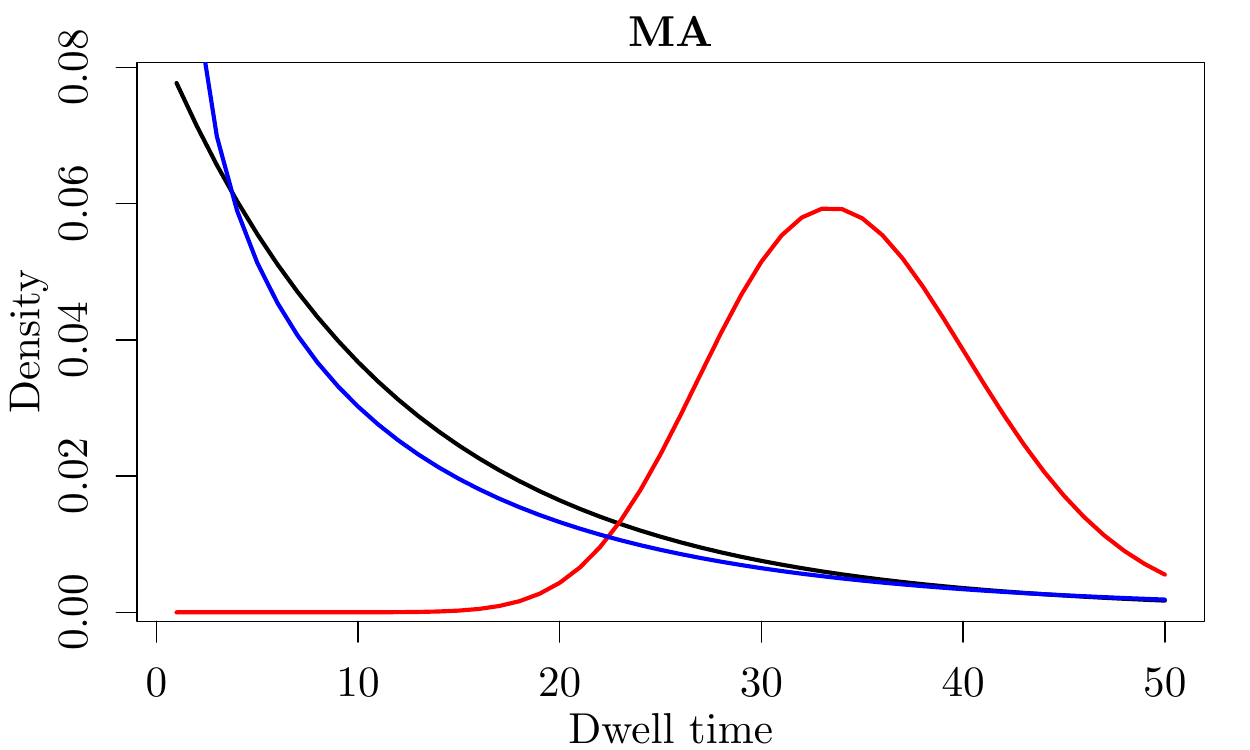}}
	\centerline{\includegraphics[width =0.49\linewidth]{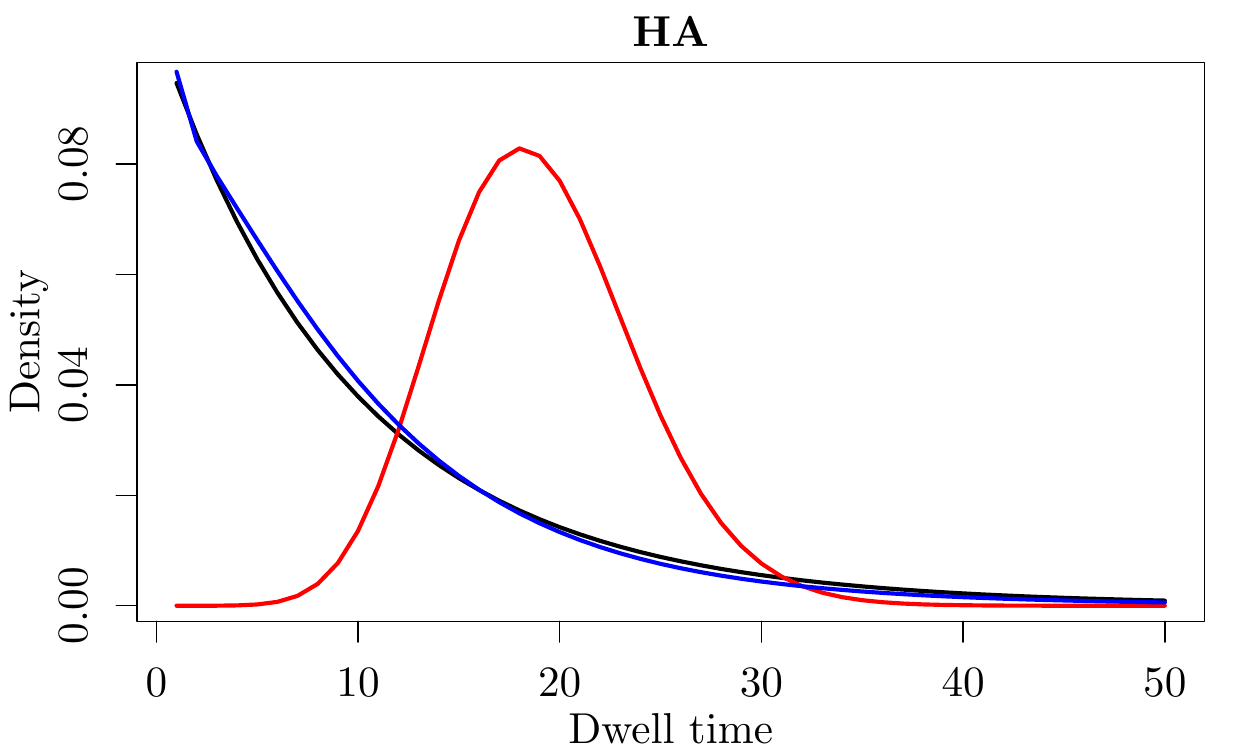}}
	\caption{Posterior predictive distribution for dwell time in \IA, \MA and \HA states under the \HMM(geometric dwell) and Poisson and negative binomial \HSMMs.}
	\label{fig:casestudy_dwells}
\end{figure}

Table 3 of the main paper provides point estimates for the parameters of the geometric (\HMM), Poisson and negative binomial dwell distributions of each of the \IA, \MA and \HA states. Figure \ref{fig:casestudy_dwells} here further plots the estimated posterior predictive distributions for the dwell length in each state. The estimated Poisson dwell distributions differ greatly from the geometric and negative binomial alternatives, in particular characterizing a much smaller variance in dwell times. The geometric and negative binomial provide more similar estimates of the dwell distribution, but notably for the \IA and \MA states the negative binomial assigns a larger probability to very short dwell times. 

\color{black}
\subsubsection{State Classification}

%\jack{To investigate the consequences of...}

We have further investigated the different state classifications provided by the optimal proposed model (using negative binomial durations) compared with the Poisson and geometric dwell distributions. Given the posterior means of the parameters of each dwell distribution, we estimated the most likely state sequence (using the Viterbi algorithm) and compared them using the confusion matrices presented in Table 1 below. 
%It seems that, from a state classification perspective, the negative binomial and geometric dwell durations perform similarly, while when choosing a Poisson durations we instead obtain significantly different results, as we have argued above. The values of the marginal likelihood, reported in Table 3 of the manuscript,  suggest that there is substantial evidence in favour of the \HSMM with negative binomial durations in comparison to a standard \HMM, and a significantly great advantage with respect to a Poisson dwell, for this applied scenario.}
%
From a state classification perspective, the negative binomial and geometric dwell durations perform similarly, while when choosing a Poisson durations we instead obtain significantly different results. This demonstrates the importance of estimating the dwell distribution in a data-driven manner, as different specifications of the dwell distribution can lead to vastly different inferential conclusions for objects of scientific interest.  The values of the marginal likelihood, reported in Table 3 of the manuscript,  suggest that there is evidence in favour of the \HSMM with negative binomial durations in comparison to a standard \HMM, and to a substantially greater extent with respect to a Poisson dwell for this applied scenario.

\begin{table}[htbp]
\centering
\begin{tabular}{clcll}
\hline \\[-0.9em] 
                     &    & \multicolumn{3}{c}{Neg-Binom}       \\
\multicolumn{1}{l}{} &    & \multicolumn{1}{l}{\IA} & \MA  & \HA   \\ \cmidrule{3-5}
Poisson              & \IA & 458                    & 5   & 0   \\
                     & \MA & 40                     & 382 & 66  \\
                     & \HA & 0                      & 1   & 198 \\[.2em] \hline
\end{tabular}
\hspace{2em}
\begin{tabular}{clcll}
\hline \\[-0.9em] 
                     &    & \multicolumn{3}{c}{Neg-Binom}       \\
\multicolumn{1}{l}{} &    & \multicolumn{1}{l}{\IA} & \MA  & \HA  \\ \cmidrule{3-5}
Geometric             & \IA & 498                    & 0   & 0   \\
                     & \MA & 0                      & 385 & 0   \\
                     & \HA & 0                      & 3   & 264 \\[.2em] \hline
\end{tabular}
\caption{State predictions summarized by the confusion matrix resulting from choosing different durations: (left) Poisson and negative binomial; (right) geometric and negative binomial.}

\end{table}

\subsection{Identifying the Periodicity}

\label{sec:periodicity}

We describe the details of our Metropolis-within-Gibbs sampler to obtain posterior samples of the frequency $\omega$, the linear basis coefficients $\bm{\beta}$, and the residual variance $\sigma^2$, under the periodic model Eq. (6.2) of the main article. This sampling scheme follows closely  the within-model move of the “segment model” introduced in  \citet{hadj2019bayesian, hadj2020spectral},  with  the  difference  that  in  this case the number  of frequencies is fixed to one. For our prior specification,  we choose a uniform prior for the frequency $ \omega \sim \text{Uniform}(0, 0.1)$ and isotropic Gaussian prior  for the vector of linear coefficients $\bm{\beta} = (\beta^{(1)}, \beta^{(2)})$ $ \sim \mathcal{\bm{N}}_{2} (\, \bm{0}, \, \sigma^2_{\beta} \, \bm{I}\, )$, where the prior variance $\sigma^2_\beta$  is fixed at 5. The prior on the residual variance $\sigma^2$ is specified as
$\text{Inverse-Gamma} \, \big(\frac{\xi_0}{2}, \frac{\tau_0}{2}\big)$, where $\xi_0 = 4$ and $\tau_0 = 1$.

For sampling the frequency, the proposal distribution is a  combination of a Normal random walk centered around the current frequency  
and a sample from the periodogram, namely
\begin{equation}
\label{mixture_proposal_freq}
q \, ( \,  \omega^{\, p}  \, | \, \omega^{\, c} \,) = \pi_{\omega} \, q_1 \, (\, \omega^{\, p}  \, | \, \omega^{\, c} \,) + (1 - \pi_{\omega} ) \, q_2 \, (\, \omega^{\, p}  \, | \, \omega^{\, c} \,)
\end{equation} where $q_1$ is defined in Eq. \eqref{q1_freq} below, $q_2$ is the density of a  Normal $\mathcal{N}\, (\omega^{\, c}, \sigma^2_{\omega})$,  $\pi_{\omega}$ is a positive value such that $ 0 \leq \pi_{\omega} \leq 1$, and the superscripts $c$ and $p$ refer to current and proposed values, respectively. For our experiments, we set $\sigma^2_{{\omega} } = 1/(25T)$ and $\pi_{\omega} = 0.1$. Eq. \eqref{mixture_proposal_freq} states that a M-H step with proposal distribution $q_1 \, (\, \omega^{\, p}  \, | \, \omega^{\, c} \,)$ \begin{equation} \label{q1_freq}
q_1 \, (\, \omega^{\, p}  \, | \, \omega^{\, c} \,) \propto \sum_{h \, = \, 0}^{T - 1} I_h \, \mathbbm{1}_{\big[ \,  h/T \, \, \leq \, \, \omega^{\, p} \, < \, \,  (h+1)/T  \, \big]  \, },
\end{equation} is performed with probability $\pi_{\omega}$, where $I_h$ is the value of the periodogram, namely the squared modulus of the Discrete Fourier transform  evaluated at frequency $h/T$
$$  I_h =  \dfrac{1}{T} \Big| \, \sum_{t=1}^{T} y_t \, \exp{\Big(-i \,  2 \pi \,  \frac{h}{T} \, \Big)}\, \Big|^{\, 2}, \quad h = 0, \dots, T-1.$$  The acceptance probability for this move is 
\begin{equation*}
\alpha  = \min \Bigg\{1, \dfrac{p \, (\omega^{\, p} \, | \, \bm{\beta},  \, \sigma^2, \, \bm{y}) }{p \, (\omega^{\,c} \, | \, \bm{\beta},  \, \sigma^2, \, \bm{y}) } \times  \dfrac{q_1 \, (\, \omega^{\, c}  \, )}{q_1 \, (\, \omega^{\, p}  \, )} \Bigg\}.
\end{equation*}
On the other hand, with probability 1 - $\pi_{\omega} $, we perform random walk M-H step with proposal distribution $q_2 \, (\, \omega^{\, p}  \, | \, \omega^{\, c} \,)$, whose density is Normal with mean $\omega^{\, c}$ and variance $\sigma^2_{{\omega} }$, i.e.
$\omega^{\, p} \, | \, \omega^{\, c} \, \sim \mathcal{N}(\,\omega^{\, c}, \, \sigma^2_{{\omega} }\,)$. This move is accepted with probability
\begin{equation*}
\alpha  = \min \Bigg\{1, \dfrac{p \, (\omega^{\,p} \, | \, \bm{\beta},  \, \sigma^2,  \, \bm{y}) }{p \, (\omega^{\, c} \, | \, \bm{\beta},  \, \sigma^2, \, \bm{y}) } \Bigg\}.
\end{equation*}

Next, we update the vector of linear coefficients $\bm{\beta}$ and the residual variance $\sigma^{\,2}$ following the usual normal Bayesian regression setting \citep{gelman2013bayesian}.
Hence,  $\bm{\beta}$ is updated in a Gibbs step  from 
\begin{equation} \label{posterior_beta}
\bm{\beta} \, \big| \,  \omega, \, \sigma^2, \,  \bm{y} \sim \bm{\mathcal{N}}_{2} \, (\, \hat{\bm{\beta}}, \, \bm{V}_{\beta}), 
\end{equation} where \begin{equation}
\begin{split}
\bm{V}_{\beta} &= \bigg( \sigma^{-2}_\beta \,  \bm{I} + \sigma^{-2} \bm{X}(\omega)^{\,'} \bm{X}(\omega) \bigg)^{-1}, \\
\hat{\bm{\beta}} &= \bm{V}_{\beta} \,  \big( \sigma^{-2} \bm{X}(\omega)^{\,'} \bm{y}  \big),
\end{split} 
\end{equation} and we denote with $\bm{X}(\omega)$ the matrix with rows given by $\bm{x}_t \, \big( \omega  \big) = [\cos(2\pi\omega t), \, \sin (2\pi \omega t)]$ for $t = 1, \dots, T$. 
Finally,  $\sigma^{\,2}$ is  drawn in a Gibbs step  directly from 
\begin{equation}
\label{inverse_gamma}
\sigma^2 \, \big| \, \bm{\beta}, \, \omega, \,  \bm{y} \sim \text{Inverse-Gamma} \, \Bigg( \, \dfrac{T + \xi_0}{2}, \, \dfrac{\tau_0 + \sum_{t=1}^{T} \Big\{ \, y_t - \bm{x}_t \, \big( \, \omega \, \big)^{\, '} \,  \bm{\beta} \, \Big\}^{2}}{2}  \Bigg).  
\end{equation}

\color{black}

%\bibliographystyle{ba}
%\bibliography{biblio}

\end{document}